\documentclass[twocolumn,english,aps,superscriptaddress,prb,floatfix,longbibliography]{revtex4-1}
\usepackage[latin9]{inputenc}
\usepackage{color}
\usepackage{babel}
\usepackage{amsmath}
\usepackage{amssymb}
\usepackage{graphicx}
\usepackage{esint}
\usepackage[unicode=true,pdfusetitle,
 bookmarks=true,bookmarksnumbered=false,bookmarksopen=false,
 breaklinks=true,pdfborder={0 0 0},pdfborderstyle={},backref=false,colorlinks=true]
 {hyperref}
\hypersetup{
 linkcolor=blue,citecolor=blue,urlcolor=blue}

\makeatletter

\usepackage{color}
\usepackage{babel}

\usepackage{color}

\newcommand{\aualyb}{\mbox{Au}_{51}\mbox{Al}_{34}\mbox{Yb}_{15}}

\makeatother

\begin{document}
\title{Conventional superconductivity in quasicrystals}
\author{Ronaldo N. Araújo}
\affiliation{Instituto de F\'{i}sica de São Carlos, Universidade de São Paulo,
C.P. 369, São Carlos, SP, 13560-970, Brazil}
\author{Eric C. Andrade}
\affiliation{Instituto de F\'{i}sica de São Carlos, Universidade de São Paulo,
C.P. 369, São Carlos, SP, 13560-970, Brazil}
\begin{abstract}
Motivated by a recent experimental observation of superconductivity
in the Al-Zn-Mg quasicrystal, we study the low-temperature behavior
of electrons moving in the quasiperiodic potential of the Ammann--Beenker
tiling in the presence of a local attraction. We employ the Bogoliubov--de
Gennes approach for approximants of different sizes and determine
the local pairing amplitude $\Delta_{i}$ as well its spatial average,
$\Delta_{0}$, the superconducting order parameter. Due to the lack
of periodicity of the octagonal tiling, the resulting superconducting
state is inhomogeneous, but we find no evidence of the superconductivity
islands, as observed in disordered systems, with $\Delta_{i}\rightarrow0$
at $T_{c}$ for all sites. In the weak-coupling regime, we find that
the superconducting order parameter depends appreciably on the approximant
size only if the Fermi energy sits at a pseudogap in the noninteracting
density of states, with $\Delta_{0}$ decreasing as the system size
increases. These results are in line with the experimental observations
for the Al-Zn-Mg quasicrystal, and they suggest that, despite their
electronic structure, quasicrystals are prone to display conventional
BCS-like superconductivity.
\end{abstract}
\date{\today}
\maketitle

\section{Introduction}

Quasicrystals display a non-periodic, yet ordered, arrangement of
atoms.\citep{shechtman84,levine84} They contain a small set of local
environments which reappear again and again, albeit not in a periodic
fashion. Their structure is not random either, since the diffraction
pattern shows sharp Bragg peaks, although their symmetry is noncrystallographic,
with the $n$-fold symmetries ($n=5,8,10,\ldots$) stemming from the
fact that these local environments occur with $n$ equiprobable orientations.
Because of this arrangement of atoms, the Bloch theorem no longer
holds and the electronic states of quasicrystals show a remarkably
rich behavior,\citep{grimm03,anu06} which includes critical states,\citep{kohmoto87,tsunetsugu91a,yuan00,tanese14,mace17}
confined states in the middle of the band,\citep{kohmoto86,arai88,rieth95}
pseudogap in the density of states,\citep{fujiwara89,fujiwara91,ishikawa17,jazbec14}
and unconventional conduction properties.\citep{pierce93,trambly94,timusk13,deLa14} 

Given their unusual electronic properties, there are several works
addressing the effects of electronic correlations in quasicrystals,
especially investigating their magnetic properties both in localized
and itinerant regimes.\citep{wessel03,vieira05a,thiem15,hartman16,koga17}
Although many interesting properties arise due of the intricate real
space arrangement of the lattice sites, some of the physical properties
inside phases with long-range order are similar to those of periodic
systems.\citep{luck93} Therefore, the experimental observation of
non-Fermi liquid behavior in the $\aualyb$ heavy-fermion quasicrystal\citep{deguchi12}
immediately prompted several theoretical studies.\citep{watanabe13,shaginyan13,andrade15,takemura15}

Superconductivity was observed in approximants,\citep{graebner87,deguchi15}
which are periodic rational approximations to the quasicrystal, shortly
after the discovery of quasicrystals. However, only recently a convincing
observation of bulk superconductivity in the Al-Zn-Mg quasicrystal
was reported.\citep{kamiya18} Reference~\onlinecite{kamiya18} finds that
the critical temperature $T_{c}$ is very low, $T_{c}\simeq0.05\mbox{ K}$,
and that $T_{c}$ is suppressed as one goes from the approximants
to the quasicrystal. Moreover, the authors show that the thermodynamic
properties can be understood within the usual BCS weak-coupling scenario.
Motivated by these experimental findings, in this paper we study the
attractive Hubbard model in a bidimensional quasicrystal. We employ
a Bogoliubov-de Gennes approach and our results provide a scenario
which is consistent with the experimental observations.

Our paper is organized as follows. In Sec. \ref{sec:Tiling-model-and},
we review the basic properties of the Ammann-Beenker tiling model
and its electronic properties. In particular, we employ the Kohn's
localization tensor to probe the spatial extent of the electronic
states. In Sec. \ref{sec:Superconductivity}, we introduce the attractive
Hubbard model and the inhomogeneous Bogoliubov-de Gennes (BdG) mean-field
theory to obtain the properties of our model inside the superconducting
phase. In Sec. \ref{sec:Discussion-and-connection}, we then compare
our results to experiments and contrast our findings with the known
results for random systems, after which we conclude the paper. We
also have two appendices. Appendix \ref{sec:Spectral-function} discusses
the spectral function of the noninteracting model, while Appendix
\ref{sec:Pairing-of-exact} introduces a complementary approach to
study the superconductivity, namely the pairing of exact eigenstates
(PoEE).

\section{\label{sec:Tiling-model-and}Tiling model and electronic properties}

\begin{figure}
\includegraphics[width=0.5\columnwidth]{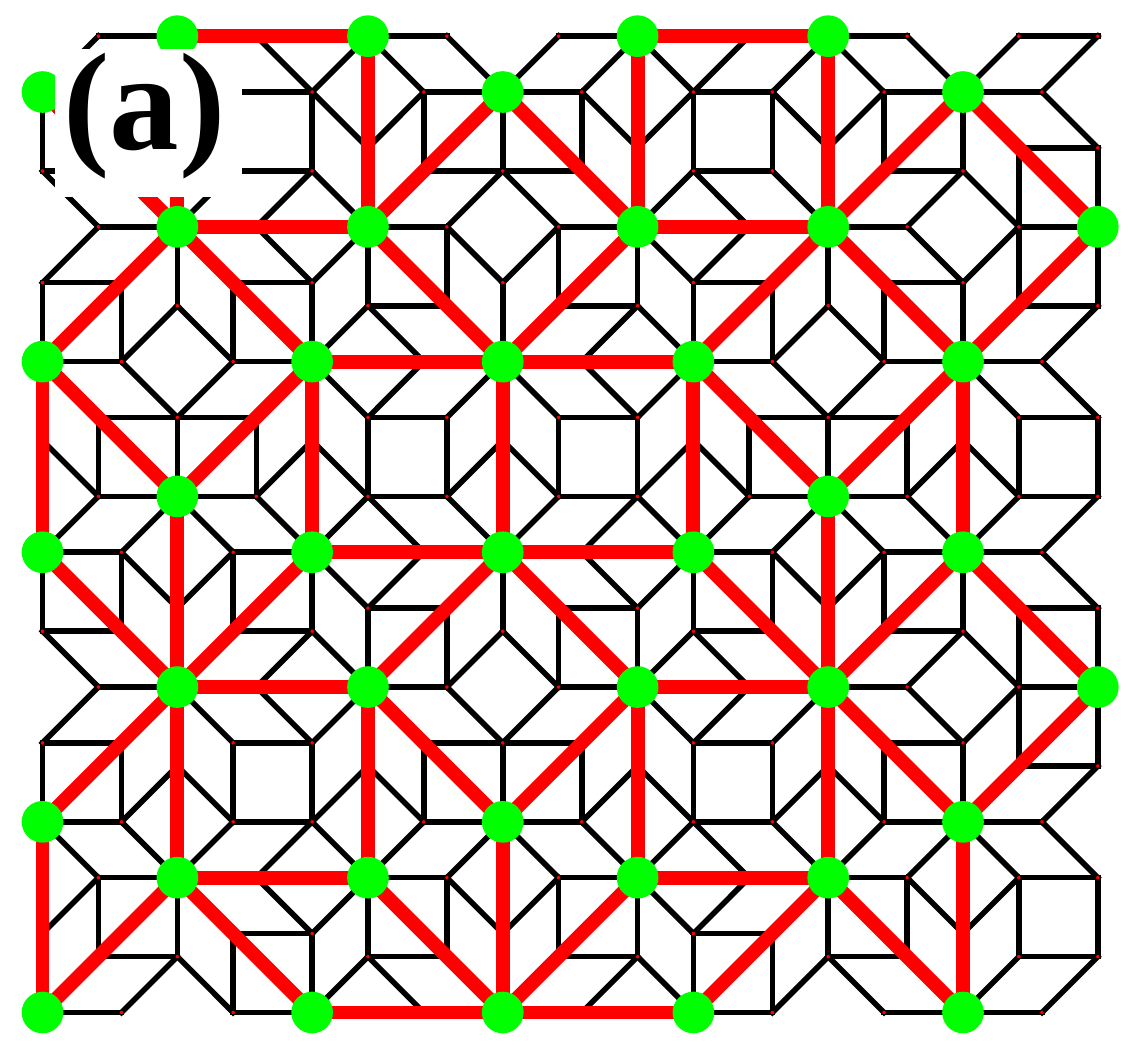}$\,$\includegraphics[width=0.46\columnwidth]{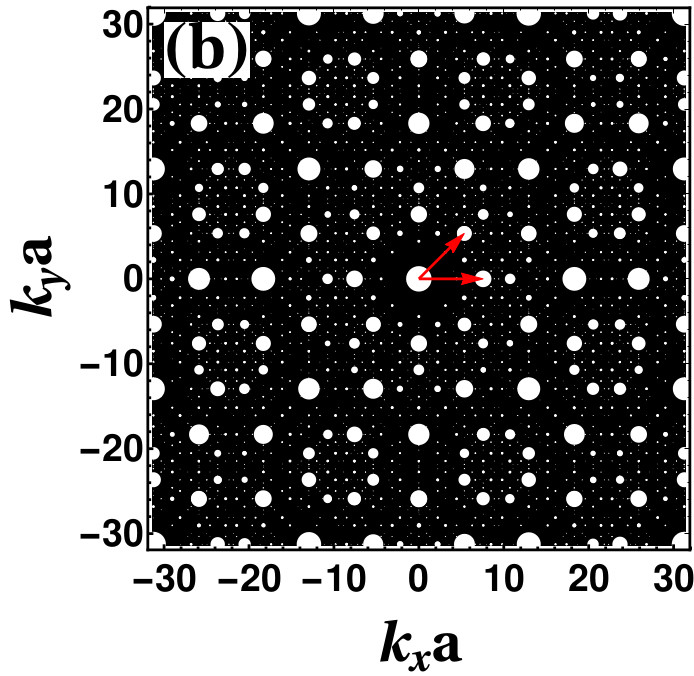}

\medskip{}

\includegraphics[width=0.48\columnwidth]{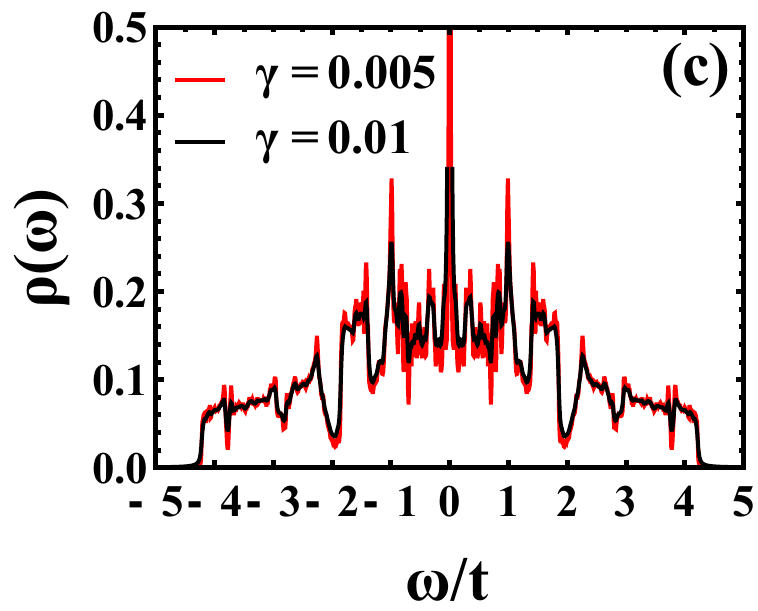}$\,$\includegraphics[width=0.55\columnwidth]{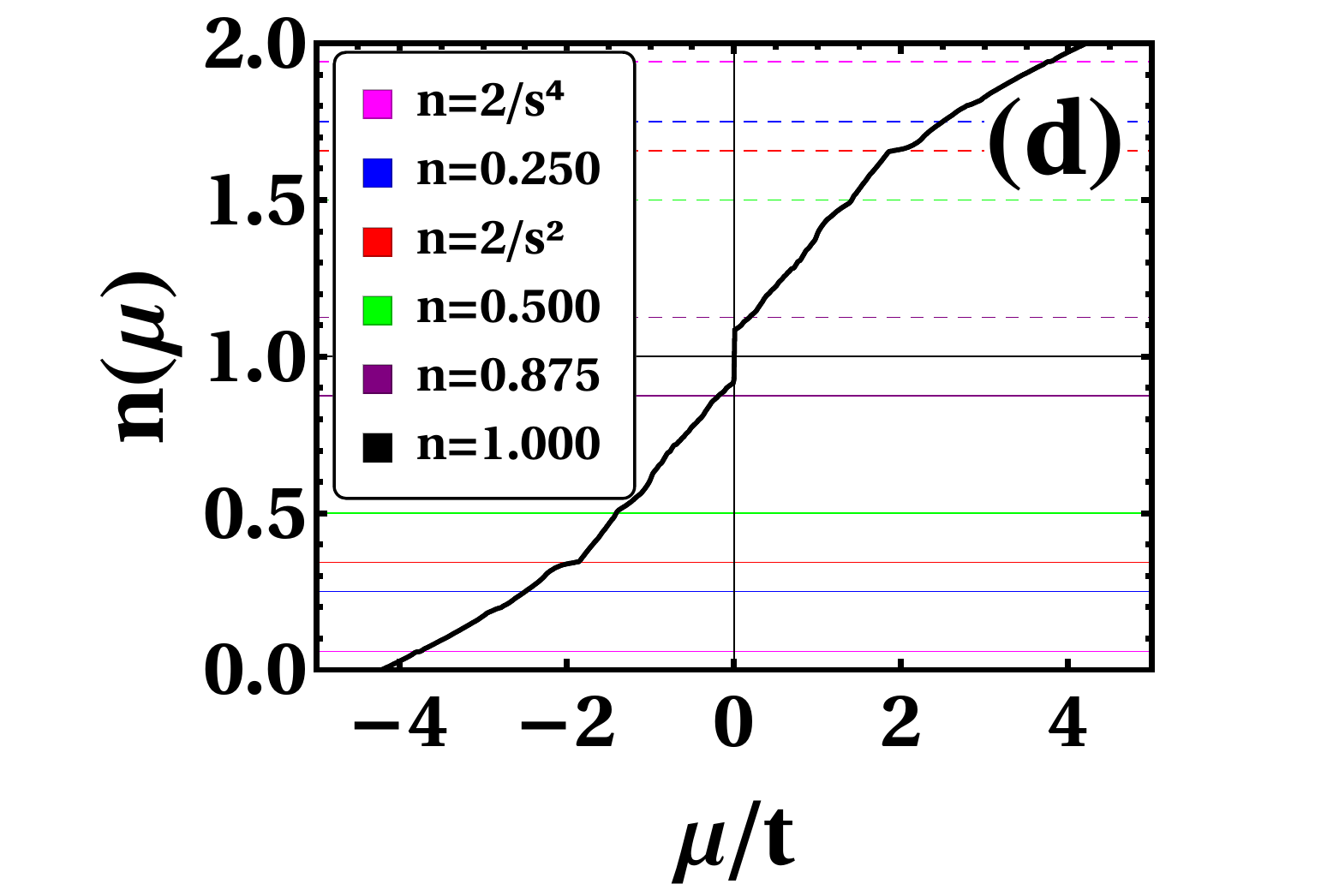}

\medskip{}

\includegraphics[width=0.5\columnwidth]{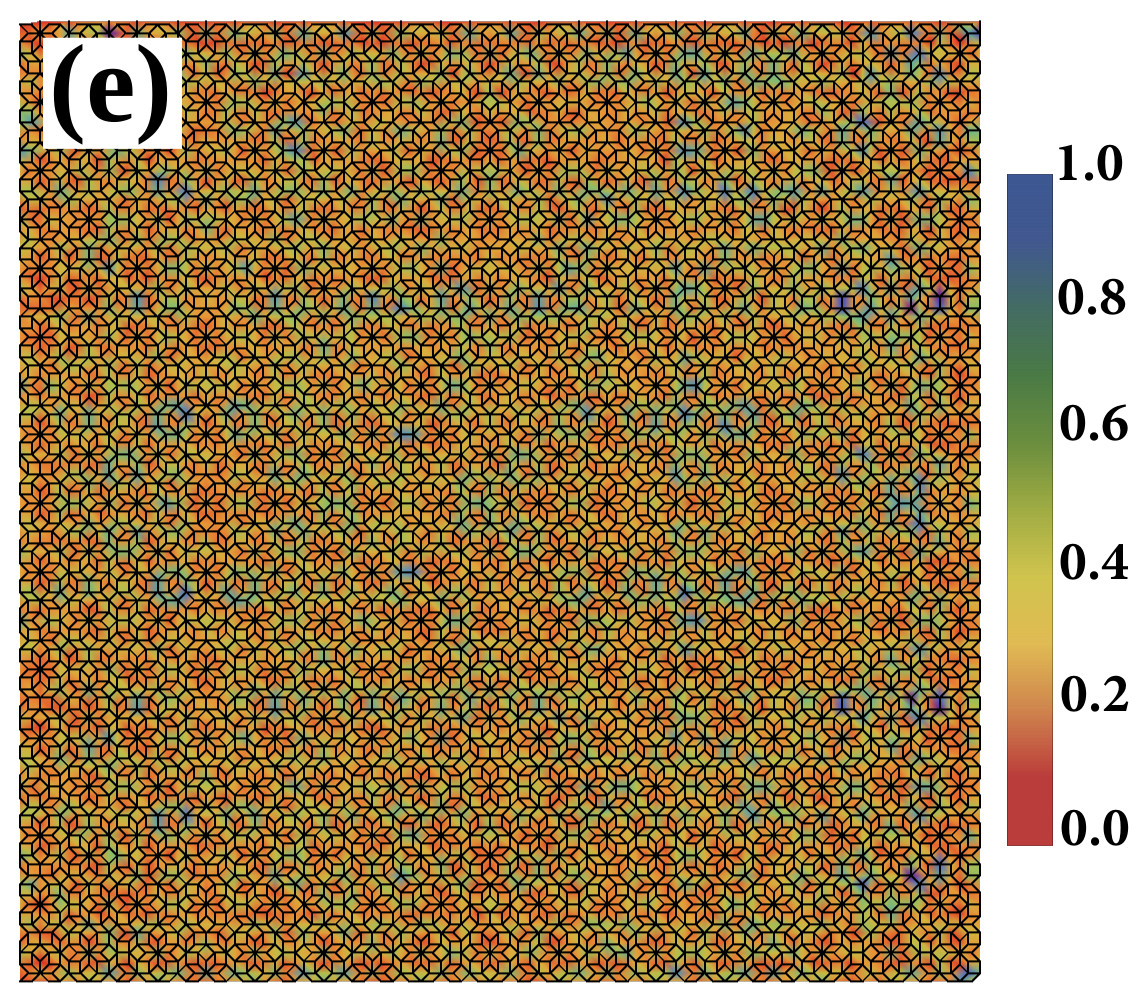}\includegraphics[width=0.5\columnwidth]{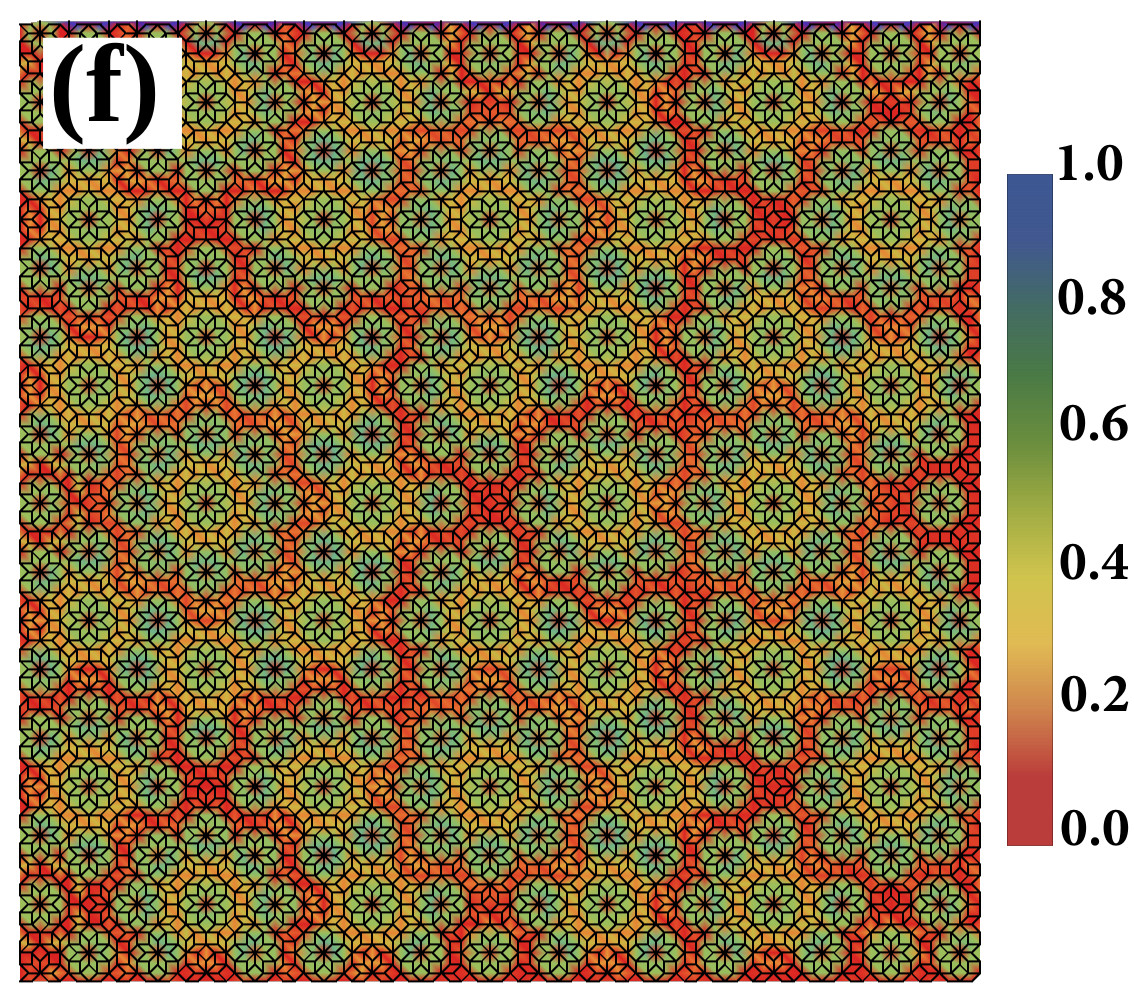}

\caption{\label{fig:tiling_def}(a) Square approximant for the perfect octagonal
tiling with $N=239$ sites. There are six local site environments
with $z=3,\ldots,8$ nearest neighbors in the bulk. Nearest-neighboring
sites are connected along the edges of the squares and the rhombuses.
Superimposed inflated $N=41$ approximant (thick lines) with its length
rescaled by the silver ratio $s=1+\sqrt{2}$. (b) X-ray structure
factor for the $N=8119$ approximant displaying the underlying eightfold
symmetry. The width of the disk indicates the intensity of the peak.
The red arrows show the vectors $\pi s\left(1,0\right)$ and $\pi s/\sqrt{2}\left(1,1\right)$,
connecting the $\Gamma$ point to two of the neighboring brighter
peaks. (c) Density of states for the $N=47321$ approximant, calculated
considering open boundary conditions and two broadening widths $\gamma$.
(d) Integrated density of states showing the electronic filling as
a function of the chemical potential. Local density of states for
the $N=8119$ approximant at the fillings: (e) $n=0.25$ and (f) $n=1.00$.}
\end{figure}

For simplicity, we consider models on quasiperiodic tilings to mimic
the geometrical properties of a quasicrystal. We report results obtained
for a $2D$ tiling, where it is easier to handle large system sizes
numerically.\citep{lcd} The $2D$ tiling we consider is the octagonal,
or Ammann-Beenker, tiling.\citep{socolar89} This tiling is composed
of two types of decorated tiles: squares and $45^{o}$ rhombuses,
which combine to create six distinct local environments with coordination
numbers $z=3,\cdots,8$, Fig.~\ref{fig:tiling_def}(a). These square
approximants are obtained by the standard method of projecting down
from a higher dimensional cubic lattice,\citep{socolar89,levine87,deneau89,benza91}
and we consider approximant of sizes $N=41$, $239$, $1393$, $8119$,
and $47321$. Even though it is easy to convince oneself on the absence
of translational invariance by examining the real-space arrangement
of the lattice sites in Fig. \ref{fig:tiling_def}(a), the eightfold
rotational symmetry (present at many different scales) becomes evident
as we go to momentum space and calculate the x-ray structure factor,
Fig. \ref{fig:tiling_def}(b). The lack of periodicity manifests itself
in reciprocal space both by the absence of a Brillouin zone and the
presence of several intense Bragg peaks. As we increase the approximant
size, more and more spots appear in the structure factor until it
becomes densely filled in the reciprocal space in the limit of the
infinite quasicrystal. Another important property of quasicrystals
is their self-similarity under inflation transformations. These are
site-decimation operations on a subset of vertices of the tiling followed
by an increase in the length scale and the reconnection of the surviving
vertices. It globally preserves the quasiperiodic structure, see Fig.
\ref{fig:tiling_def}(a), and the infinite quasicrystal is invariant
under such transformation.\citep{deneau89,anu00}

The presence of diffraction spots of widely differing intensities
has important consequences on the electronic properties of the quasicrystal.
In a periodic system, if the Fermi wave vector $\mathbf{k}_{F}$ satisfies
$2\mathbf{k}_{F}=\mathbf{H}$, where $\mathbf{H}$ is a reciprocal
lattice vector, a band gap is expected to emerge. In a quasicrystal,
because the structure factor is densely filled, this condition is
easily met and then we expect the brighter peaks to lead to strong
scattering of conduction electrons, giving rise to spikes in the density of states (DOS)
[see also Fig. \ref{fig:arpes}(b)].\citep{smith87,zijlstra00} The
scattering due to the remaining peaks, while weaker, results in wave functions
which show fluctuations at all length scales. The Fibonacci chain,
a one-dimensional quasicrystal, provides an example of such wavefunctions,\citep{kohmoto87}
often referred to as \emph{critical},\citep{kohmoto87,tsunetsugu91a,yuan00,grimm03,anu06,mace17}
in analogy with those found at the Anderson metal-insulator transition.\citep{richardella10,rodriguez10,rodriguez11}

As a minimal model to describe the electronic properties of quasicrystals,
we study a nearest-neighbor tight-binding Hamiltonian in the Ammann-Beenker
tiling,

\begin{eqnarray}
\mathcal{H}_{0} & = & -t\sum_{\left\langle ij\right\rangle ,\sigma}\left(c_{i\sigma}^{\dagger}c_{j\sigma}+c_{j\sigma}^{\dagger}c_{i\sigma}\right),\label{eq:h0}
\end{eqnarray}
where $c_{i\sigma}^{\dagger}\left(c_{i\sigma}\right)$ is the creation
(annihilation) operator of an electron at site $i$ with spin $\sigma$
and $t$ is the hopping amplitude between sites $i$ and $j$. In
the following, energies are measured in units of $t$. In our calculation,
we consider open boundary conditions because: (i) it preserves the
particle-hole symmetry of the tiling; and (ii) the finite size effects
are comparable to those of periodic boundary conditions due to the
quasiperiodic arrangement of the different local environments.\citep{hartman16}

The resulting DOS $\rho\left(\omega\right)=\left(1/N\right)\sum_{\nu}\delta\left(\omega-\varepsilon_{\nu}\right)$,
where $\varepsilon_{\nu}$ are the eigenenergies of $\mathcal{H}_{0}$
in Eq. \eqref{eq:h0}, is shown in Fig. \ref{fig:tiling_def}(c) (we replace
each delta function by a Lorentzian of width $\gamma$). As anticipated,
$\rho\left(\omega\right)$ displays a strong energy dependence with
several spikes, which are largely independent of the broadening $\gamma$.
The large peak at $\omega=0$ can be traced to families of strictly
localized states, a consequence of the local topology of the octagonal
tiling.\citep{grimm03,rieth95} The integrated density of states $n\left(\mu\right)=\int_{-\infty}^{\mu}d\omega\rho\left(\omega\right)$,
$\mu$ is the chemical potential, is shown in Fig. \ref{fig:tiling_def}(d).
Besides the discontinuity close to $\mu=0$, corresponding to the
peak in $\rho\left(0\right)$, $n\left(\mu\right)$ also shows a kink
at the filling $2/s^{2}\approx0.34315$, where $s=1+\sqrt{2}$ is
the silver ratio. This is analogous to the case of the Fibonacci chain,
where plateaus in $n\left(\mu\right)$ appear at $2/g^{n}$, where
$n$ is an integer and $g$ is the golden ratio.\citep{tanese14}
A plateau in $n\left(\mu\right)$ corresponds to a gap in the single-particle
spectrum of Eq. \ref{eq:h0}. However, conversely to the one-dimensional
case, the DOS in the Ammann-Beenker tiling has at most
a pseudogap close to $\omega\approx-1.9t$, corresponding to the filling
$2/s^{2}$, and thus we observe only a kink.

A pseudogap at the Fermi level assists in the stabilization of the
quasiperiodic structure via the Hume--Rothery mechanism, and it is
indeed predicted and observed in several quasicrystals.\citep{smith87,fujiwara91,trambly94,stadinik97,rotengerg00,rogalev15,jazbec14}

The unambiguous existence of a pseudogap is hindered due to the finite
broadening $\gamma$ employed in the numerical calculation of $\rho\left(\omega\right)$.
Therefore, we now probe the spatial extent of the wave function, especially
close to the filling $2/s^{2}$. First, we compute the inverse participation
ratio (IPR),
\begin{eqnarray}
IPR_{\nu} & = & \sum_{i}\left|\psi_{\nu}\left(i\right)\right|^{4},\label{eq:ipr}
\end{eqnarray}
where $\psi_{\nu}$ is an eigenstate of $\mathcal{H}_{0}$ with eigenenergy
$\varepsilon_{\nu}$. The scaling of the IPR with the system size
is related to the spatial structure of the single-particle electronic
states. If we write $IPR_{\nu}\propto N^{-\beta}$, then $\beta=1$
for extended and $\beta=0$ for exponentially localized states. In
a quasicrystal, we expect $0\le\beta\le1$, due to the multifractal
character of the eigenstates.\citep{grimm03,chhabra89,richardella10,rodriguez10}
In Fig. \ref{eq:ipr} we calculate $IPR_{\nu}$ at different positions
in the band. For most fillings, we obtain $\beta\approx0.90$, a value
similar to the one observed in the Penrose tiling.\citep{grimm03}
At the band center, we get $\beta\approx1$, a value one expects for
extended states (at half filling, we have flat bandlike structures
coexisting with dispersive ones, see Appendix \ref{sec:Spectral-function}).
For $n=2/s^{2}$, we have $\beta=0.55\left(6\right)$. Although smaller
than the values at the other fillings, this value does not indicate
that this particular state is localized and it seems inconsistent
with the presence of a pseudogap.

\begin{figure}
\includegraphics[width=0.5\columnwidth]{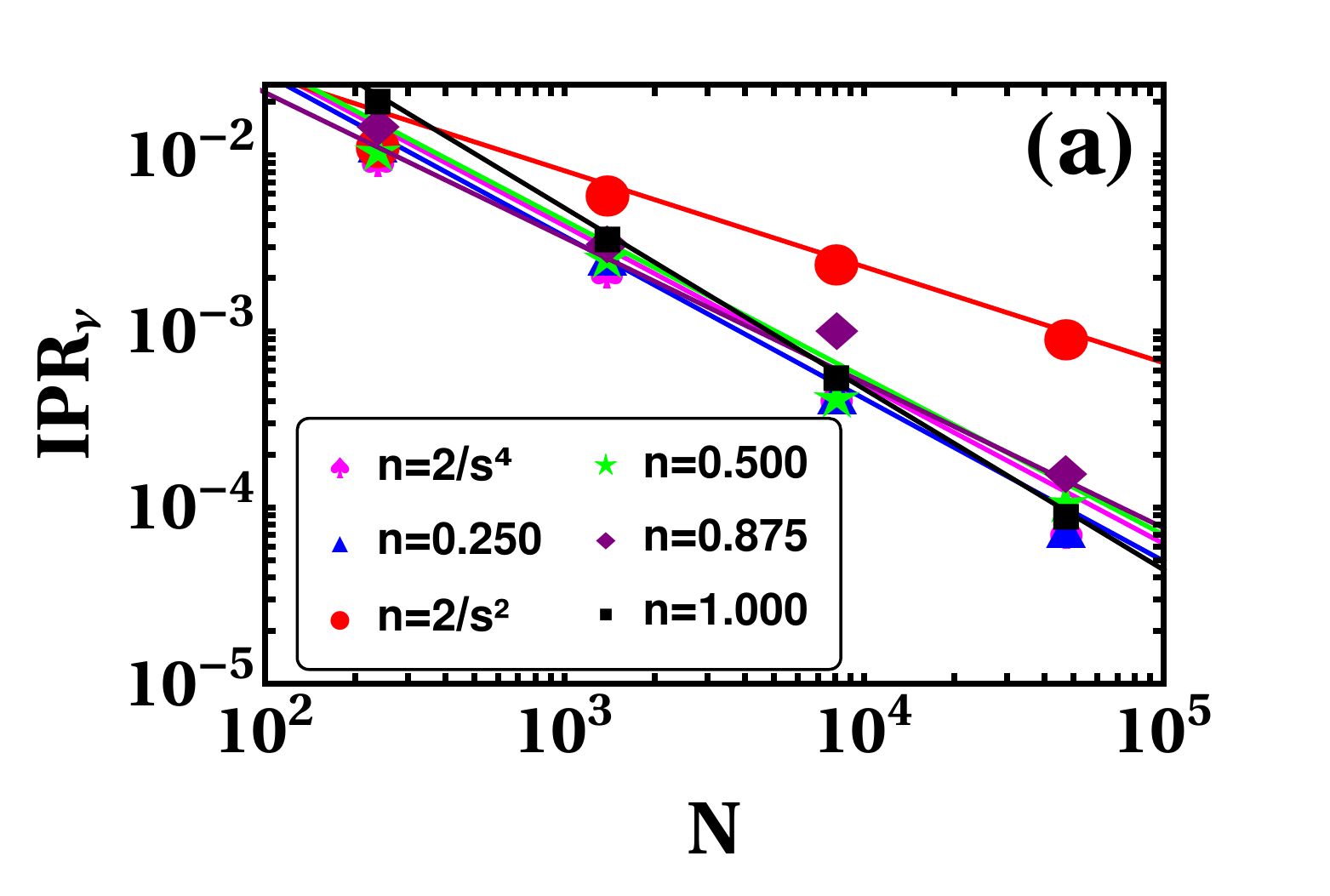}$\,$\includegraphics[width=0.5\columnwidth]{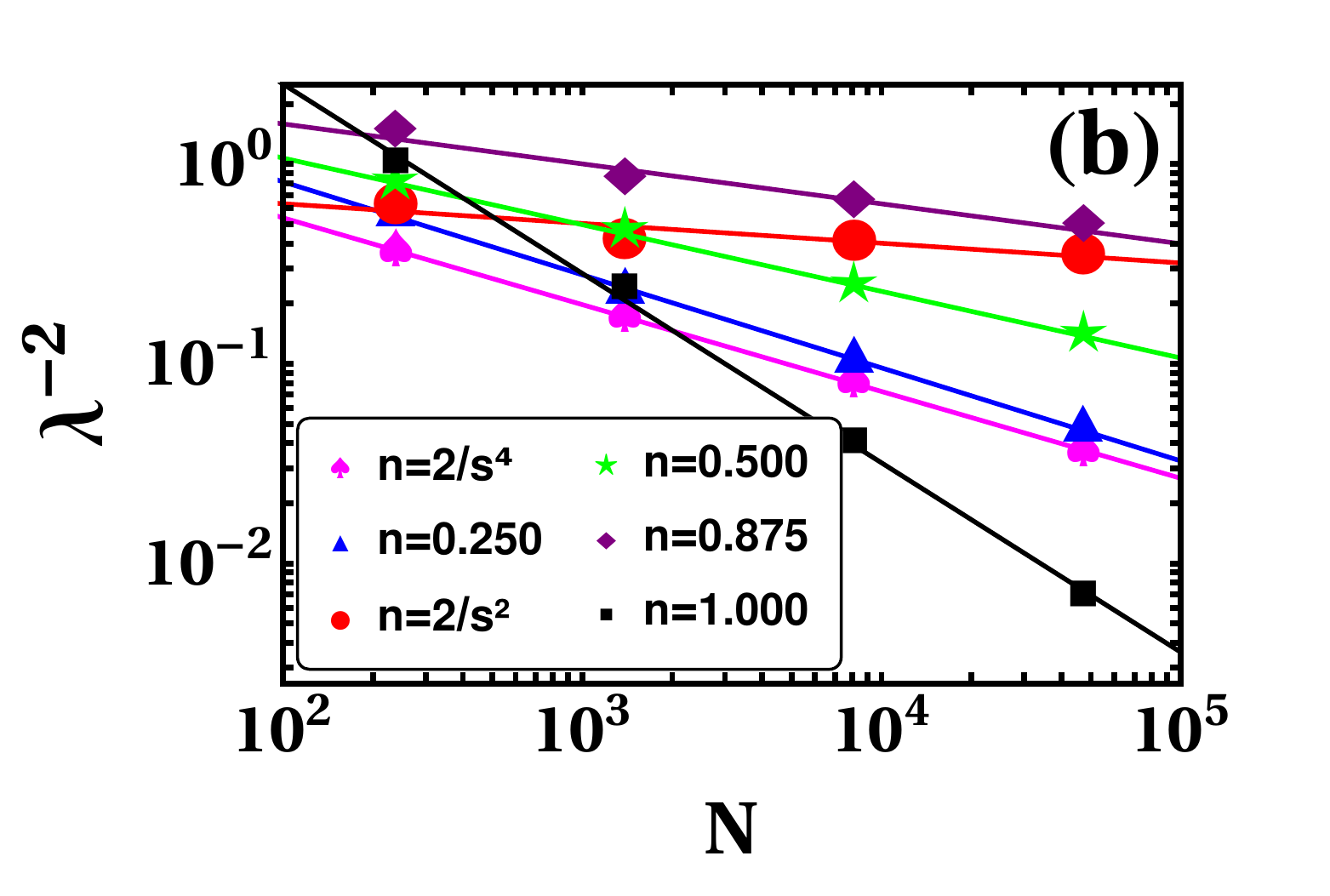}

\caption{\label{fig:ipr_kohn}(a) Inverse participation ratio $IPR_{\nu}$
as a function of the approximant size $N$ for different fillings
$n$ on a log-log plot. We fit $IPR_{\nu}\propto N^{-\beta}$. (b)
Same as (a) for the Kohn's localization tensor. We fit $\lambda^{-2}\propto N^{-\alpha}$.}
\end{figure}

While the IPR is a very useful tool in the context of disordered systems,
it may not be able to capture all the subtleties of quasicrystalline
electronic states. Indeed, a recent study of the one-dimensional Fibonacci
chain showed that the IPR is unable to capture the expected insulating
behavior inside the band minigaps.\citep{varma16} Following Ref.~\onlinecite{varma16},
we then decided to study the scaling behavior of the Kohn's localization
tensor\citep{resta99,resta11,souza00}
\begin{equation}
\lambda_{\gamma\delta}=\frac{1}{N}\sum_{i,j}\left(\mathbf{r}_{i}-\mathbf{r}_{j}\right)_{\gamma}\left(\mathbf{r}_{i}-\mathbf{r}_{j}\right)_{\delta}\left|P\left(i,\,j\right)\right|^{2}.\label{eq:kohn_def}
\end{equation}
Here $\mathbf{r}_{i}$ is the position of site $i$ inside the approximant,
$\gamma,\,\delta$ correspond to the spatial directions $x$ and $y$,
and $P\left(i,\,j\right)=\sum_{\nu}\psi_{\nu}\left(i\right)\psi_{\nu}^{\star}\left(j\right)$,
with $\varepsilon_{\nu}\le\mu$, is the one-particle density matrix
for a Slater determinant. Therefore, the localization tensor takes
into account all states up to the chemical potential and not only
a single-particle state. Because time-reversal symmetry is preserved
in the problem, the transverse terms vanish identically: $\lambda_{xy}=\lambda_{yx}=0$.
Moreover, we have $\lambda_{xx}=\lambda_{yy}=\lambda$, so we drop
the spatial subscripts henceforth. The scaling of length $\lambda$
with the approximant size then determines if the system is a metal
or an insulator. In a metal, we expect $\lambda^{2}$ to diverge with
$N$, whereas for an insulator we expect $\lambda^{2}$ to saturate
to a constant.\citep{varma16} If we write $\lambda^{-2}\propto N^{-\alpha}$,
we then expect $0\le\alpha\le1$. The results for the scaling of Kohn's
localization tensor are displayed in Fig. \ref{fig:ipr_kohn}(b),
where it is clear that its dependence with the band filling is indeed
more pronounced as compared to the IPR. For arbitrary filling, the
states have an extendedlike nature, particularly at the band center
where we have $\alpha\approx1$. For the filling $n=2/s^{2}$, however,
$\lambda^{-2}$ is weakly size dependent, $\alpha\approx0.1$, suggesting
a localized-like nature for this state, consistent with the presence
of a pseudogap.

Overall, we find that the Kohn's localization tensor has a stronger
dependence with the band filling and it is better suited to decide
whether the electronic states are conducting or insulating.\citep{varma16}
We stress, however, that $\lambda$ is not simply related to the spatial
extent of the single-particle eigenstates, and a more detailed characterization
of the multifractal character of the eigenstates is not straightforward
within this formalism.

\section{\label{sec:Superconductivity}Superconductivity}

After discussing the electronic properties of the octagonal tiling,
we now move to the main topic of this paper which is the study of superconductivity.
We describe an $s$-wave superconductor using the attractive Hubbard
Hamiltonian, 
\begin{equation}
\mathcal{H}=\mathcal{H}_{0}-U\sum_{i}n_{i\uparrow}n_{i\downarrow},\label{eq:hubbard}
\end{equation}
where $\mathcal{H}_{0}$ is given by Eq. \eqref{eq:h0}, $U>0$ is
the uniform on-site pairing attraction, and $n_{i\sigma}=c_{i\sigma}^{\dagger}c_{i\sigma}$
is the number operator. This model naturally neglects the effects
of Coulomb interaction, as well as the nontrivial phonon spectrum
of quasicrystals.\citep{luck86b,los93,quilichini97,deboissieu12,brown18}
However, we feel that it is a useful exercise to understand the physics
of this simple model, where we can contrast our results with a similar
investigation in the Penrose tiling.\citep{sakai17} Moreover, we
will show that it provides a useful starting point to understand the
recently reported quasicrystal superconductivity.\citep{kamiya18}

We study the model in Eq. \eqref{eq:hubbard} for different values
of the pairing attraction $U$ and filling $n$. Because the DOS in
a quasicrystal is strongly energy dependent, Fig. \ref{fig:tiling_def}(c),
one could expect, in principle, strong filling dependence of the results.
Nevertheless, as we discussed in the previous section, the behavior
of the electronic states is qualitatively the same for all fillings,
i.e., metalliclike, except at the special filling of $2/s^{2}$ where
we observe an insulating behavior due to a pseudogap in the DOS.

To solve the Hamiltonian \eqref{eq:hubbard}, we employ the BdG approach,
following the works in Refs.~\onlinecite{ghosal98,ghosal01}, to write
the mean-field Hamiltonian: 
\begin{align}
\mathcal{H}_{BdG} & =-t\sum_{\left\langle ij\right\rangle ,\sigma}\left(c_{i\sigma}^{\dagger}c_{j\sigma}+c_{j\sigma}^{\dagger}c_{i\sigma}\right)-\sum_{i,\sigma}\tilde{\mu}_{i}n_{i\sigma}\nonumber \\
 & +\sum_{i}\left(\Delta_{i}c_{i\uparrow}^{\dagger}c_{i\downarrow}^{\dagger}+\Delta_{i}^{\star}c_{i\downarrow}c_{i\uparrow}\right).\label{eq:hbdg}
\end{align}
The local pairing amplitude $\Delta_{i}$ and the local density $n_{i}$
are determined via the self-consistent equations,
\begin{equation}
\Delta_{i}=-U\left\langle c_{i\downarrow}c_{i\uparrow}\right\rangle ,\,\left\langle n_{i}\right\rangle =\sum_{\sigma}\left\langle c_{i\sigma}^{\dagger}c_{i\sigma}\right\rangle ,\label{eq:self_bdg}
\end{equation}
where the thermal averages are taken considering the eigenstates and
eigenenergies of Eq. \eqref{eq:hbdg}, which we determine via a numerical
Bogoliubov transformation. We also introduce an effective chemical
potential to incorporate a site-depend Hartree shift: $\tilde{\mu}_{i}=\mu+U\left\langle n_{i}\right\rangle /2$,
building thus the most general mean-field theory for an inhomogeneous
$s$-wave superconductor.\citep{vlad_book} We remark that this mean-field
theory keeps only the amplitude fluctuations of $\Delta_{i}$ and
should only be valid at weak-coupling.

\begin{figure}
\includegraphics[width=0.5\columnwidth]{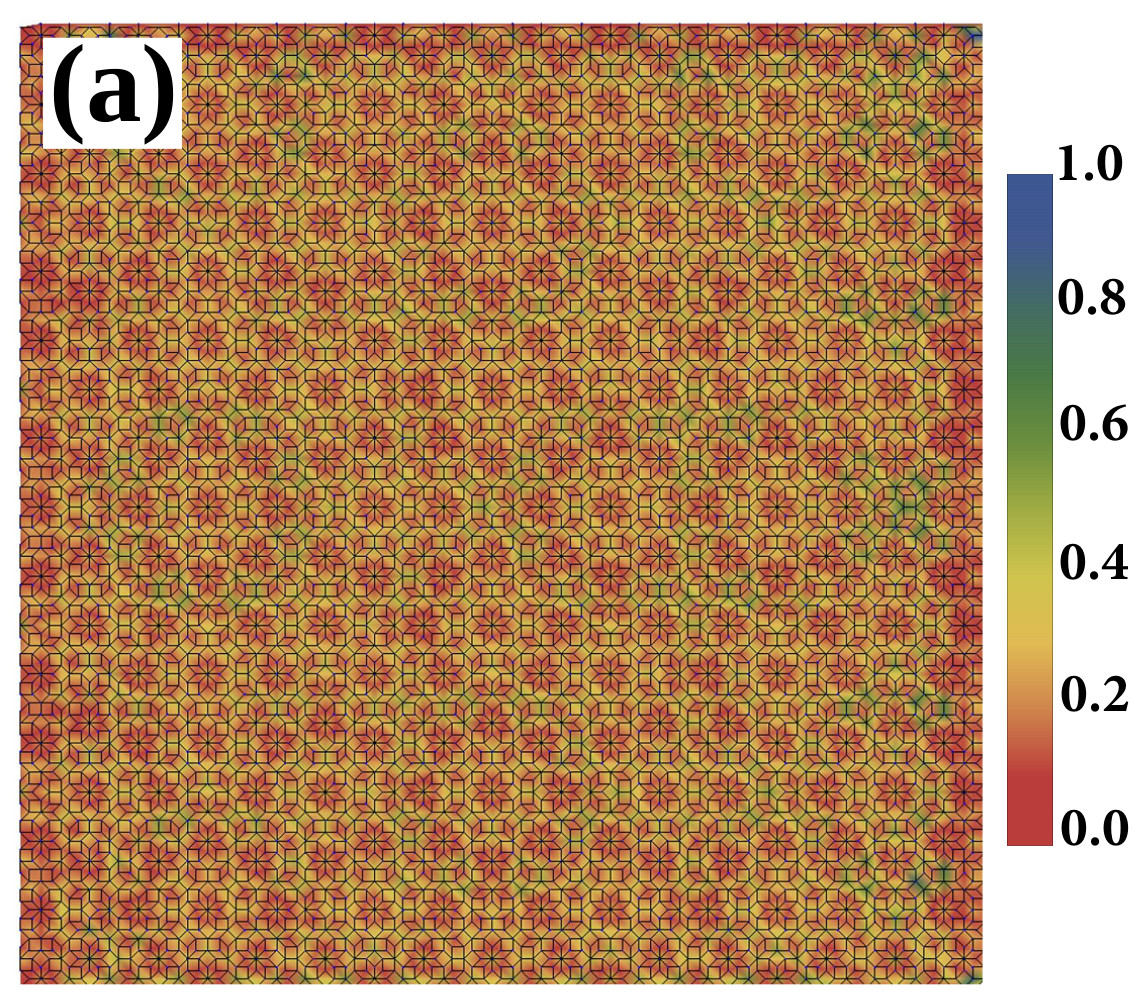}\includegraphics[width=0.5\columnwidth]{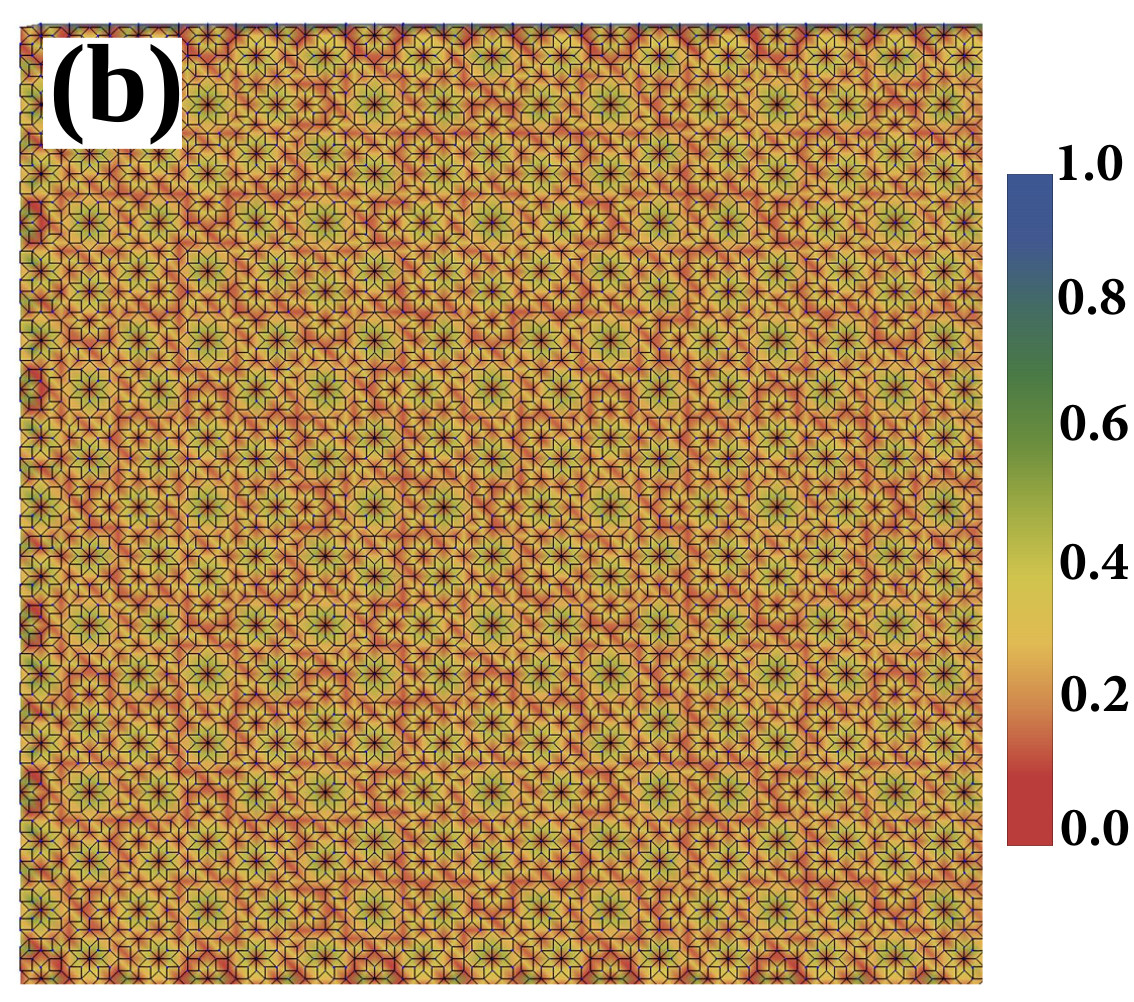}

$\,$

\includegraphics[width=0.47\columnwidth]{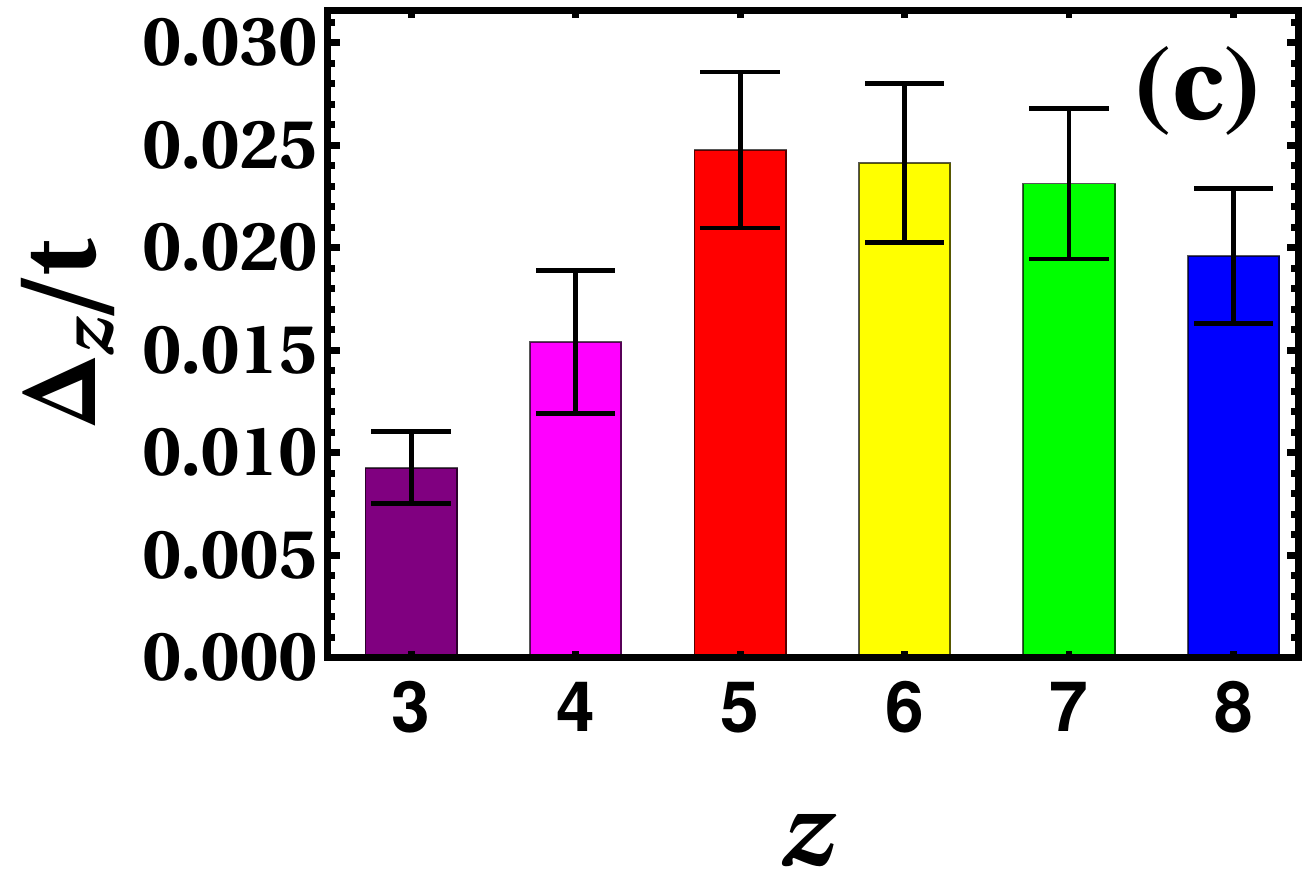}$\,$\includegraphics[width=0.47\columnwidth]{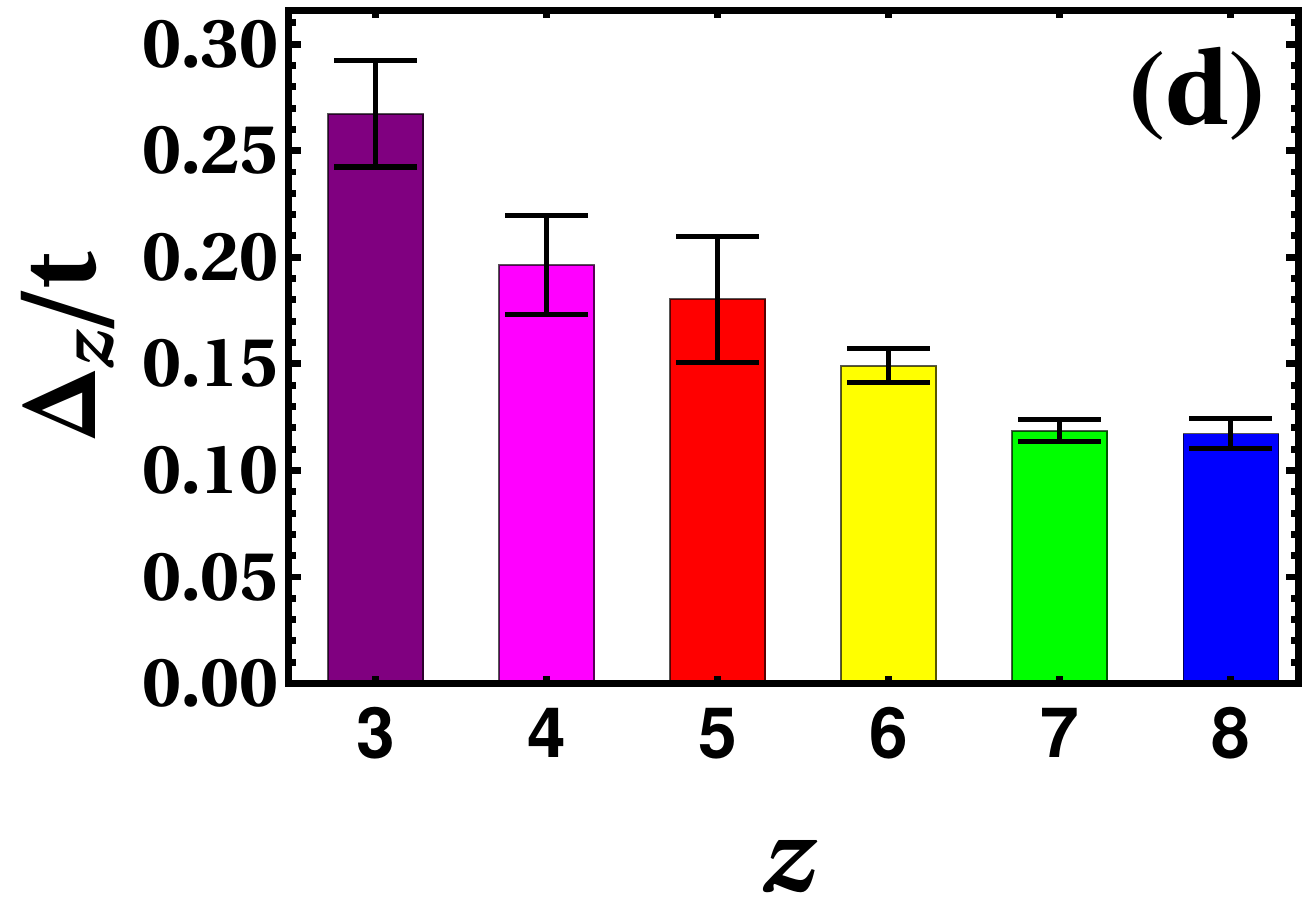}

$\,$

\includegraphics[width=0.47\columnwidth]{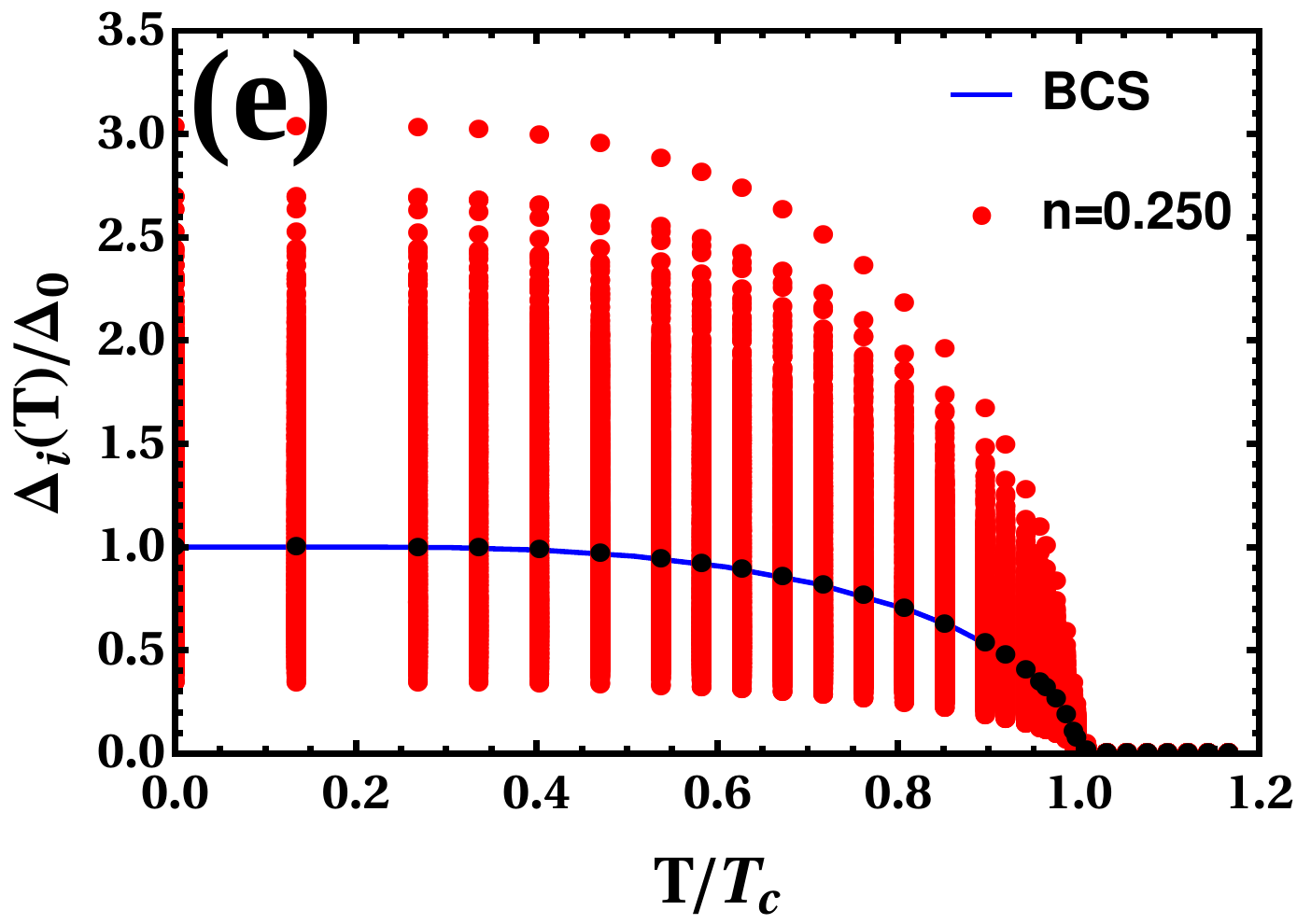}$\,$\includegraphics[width=0.47\columnwidth]{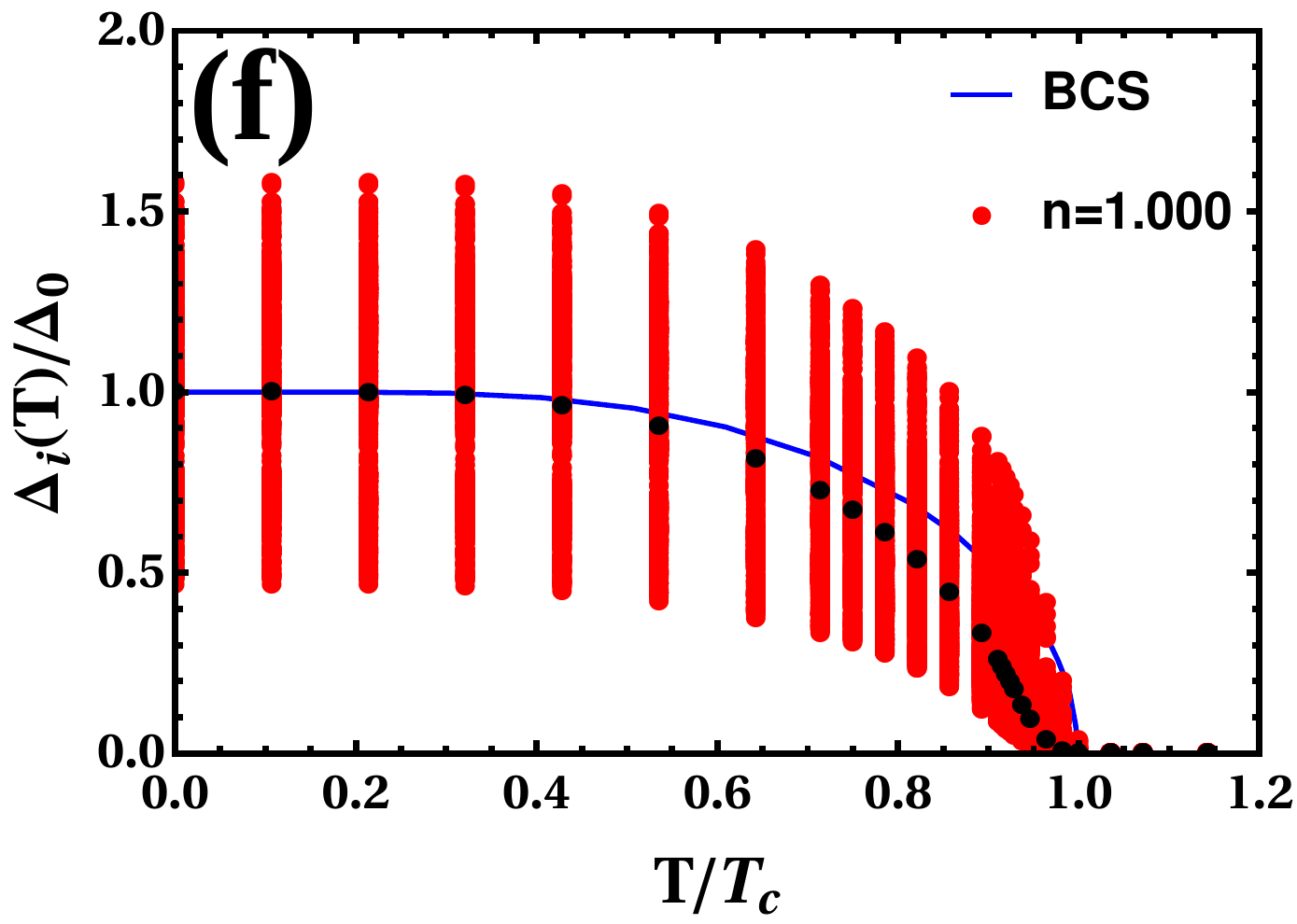}

\caption{\label{fig:delta_i} Color plots of local pairing amplitude $\Delta_{i}$,
normalized by its maximum value: (a) $n=0.25$ and (b) $n=1.00$.
Average value of the local pairing amplitude for a given local environment
characterized by the coordination number $z$, $\Delta_{z}$. The
height of the bars give the average value and the black lines the
standard deviation: (c) $n=0.25$ and (d) $n=1.00$. Here we considered
$T=0$, the approximant with $N=8119$ sites and $U=1.5t$. Temperature
dependence, rescaled by $T_{c}$, of the the local pairing amplitude
$\Delta_{i}$, divided by $\Delta_{0}$, for the $N=1393$ approximant
and two different fillings: (e) $n=0.25$ $\left[T_{c}/t=0.011,\,\Delta_{0}=0.015\left(6\right)\right]$
and (f) $n=1.00$ $\left[T_{c}/t=0.13,\,\Delta_{0}=0.22\left(5\right)\right]$.
We compare the results with BCS theory (blue curve). The black dots
represent the average value of $\Delta_{i}$, whereas the red dots
show all values of $\Delta_{i}$ for the tiling.}
\end{figure}

We solve the self-consistency Eqs. \eqref{eq:self_bdg} on finite
approximants with $N$ sites, open boundary conditions, and fixed
electronic filling $n=\sum_{i}n_{i}/N$. We consider $U\ge1.5t$ because
smaller values of $U$ generate very large coherence lengths and are
harder to simulate. To solve Eqs. \eqref{eq:self_bdg}, we start with
an initial guess for the local density $n_{i}$ and the pairing amplitude
$\Delta_{i}$, and we iterate the procedure until convergence is achieved
on all sites. We then adjust the chemical potential $\mu$ to target
the desired filling $n$. Notice, therefore, that we have two self-consistency
loops, making the whole procedure quite demanding numerically. To
complement this procedure, and to access larger system sizes, we also
implement the method of PoEE\citep{ghosal98,ghosal01} (see Appendix
\ref{sec:Pairing-of-exact}).

We start by showing the spatial distribution of the local pairing
amplitude $\Delta_{i}$ at $T=0$ in Figs. \ref{fig:delta_i} (a) and
\ref{fig:delta_i} (b). For small values of the pairing attraction $U$, the spatial
pattern of $\Delta_{i}$ roughly follows that of the local density
of states,\citep{collins17} see Figs. \ref{fig:tiling_def} (e) and
\ref{fig:tiling_def} (f), and it is essentially determined by the local environment of
a given site $i$. At low fillings, the sites with a larger coordination
number $z$ show larger values of $\Delta_{i}$, whereas for $n\rightarrow1$
sites with smaller $z$ are the ones with larger $\Delta_{i}$, Figs.
\ref{fig:delta_i} (c) and \ref{fig:delta_i} (d). Importantly, we do not observe the
formation of superconducting islands as in disordered superconductors.\citep{ghosal98,ghosal01,bouadim11,nandkishore13,potirniche14,dodaro18}
This is not unexpected because these islands occur in regions where
the random disorder potential is unusually small, corresponding thus
to rare regions,\citep{amd09,hoyos14} a situation which cannot take
place in the presence of a deterministic quasiperiodic potential obeying
inflation rules. Because the distribution of $\Delta_{i}$ consists
essentially of six delta peaks, each one associated to a local environment,
and is neither broad nor shows weight at $\Delta_{i}\approx0$, we
also conclude that the system is far away from a possible quasiperiodicity-induced quantum phase transition.

In smaller approximants, $N\le1393$, we are able to solve the BdG
solutions at finite temperatures, Figs. \ref{fig:delta_i} (e) and
\ref{fig:delta_i} (f). Even though $\Delta_{i}$ is spatially inhomogeneous, we find
that the superconducting phase transition takes place at all approximant
sites at once within our numerical precision. In a translational invariant
system, we naturally expect all $\Delta_{i}$'s to vanish concomitantly
at $T_{c}$. For the octagonal tiling, we believe that its self-similarity
under inflation transformations, see Fig. \ref{fig:tiling_def}(a),
forces all $\Delta_{i}$'s to vanish simultaneously at $T_{c}$. To
see this, suppose we start with a subset of sites for which $\Delta_{i}=0$
inside the superconducting phase. Now we successively apply the inflation
transformations (remember also that $\Delta_{i}$ is essentially determined
by its local environment). Because these transformations leave the
infinite system invariant due to their self-similarity, we would then
be able to eventually move the entire system to the normal phase.
This argument also highlights that the existence of rare regions is
not possible, thus precluding both the existence of the superconducting
islands at $T=0$, as discussed before, and the presence of a thermal
Griffiths phase close to $T_{c}$.

\begin{figure}
\includegraphics[width=0.5\columnwidth]{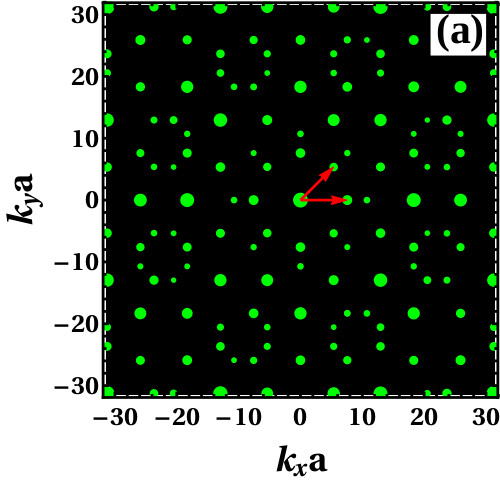}\includegraphics[width=0.5\columnwidth]{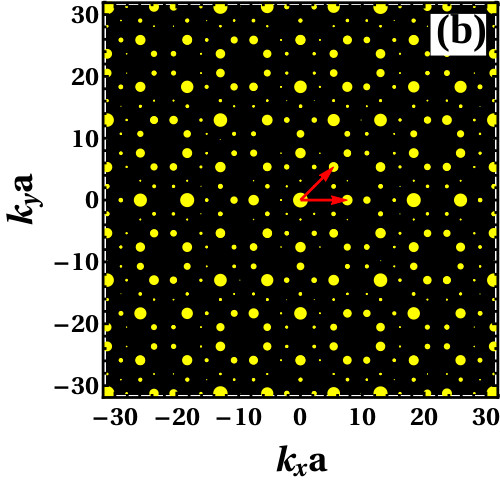}

\includegraphics[width=0.5\columnwidth]{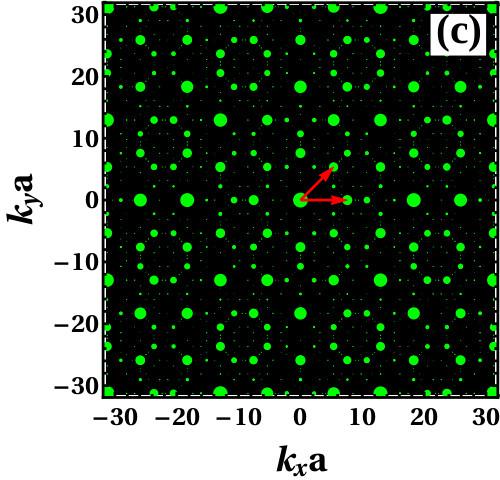}\includegraphics[width=0.5\columnwidth]{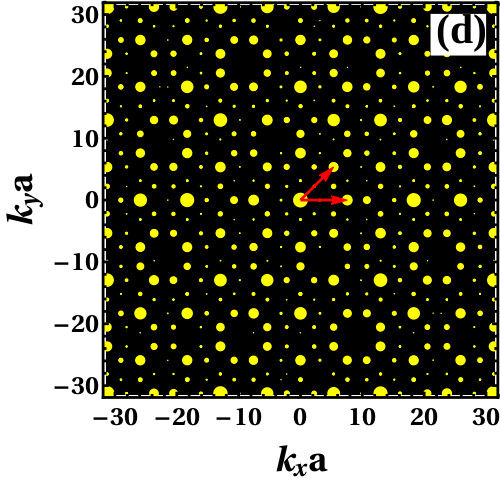}

\caption{\label{fig:fourier} Fourier transform of the local pairing amplitude,
(a) and (c), and the local electronic density $n_{i}$, (b) and (d),
for $n=2/s^{2}$ with $U=1.5t$, (a) and (b), and $U=6.0t$, (c) and
(d). The width of the disk indicates the intensity of the peak. The
red arrows are the same as in Fig. \ref{fig:tiling_def}(b). Here
we considered $N=8119$.}
\end{figure}

To explore the role of the coherence length $\xi$ --- as defined
by the spatial decay of the sample averaged correlation function $\left\langle \Delta_{i}\Delta_{j}\right\rangle $
--- we study the Fourier transform of both $\Delta_{i}$ and $n_{i}$,
see Fig. \ref{fig:fourier}. The Fourier transform of $\Delta_{i}$
shows the expected eightfold structure, as in Fig. \ref{fig:tiling_def}(b),
but there are several missing peaks which we link to the presence
of a coherence length $\xi$ ($\xi$ is the largest, circa $10$ lattice
spacing, at the pseudogap $n=2/s^{2}$). As we increase the local
attraction $U$, the Cooper pairs become more and more local, resulting
in the suppression of $\xi$ and in the observation of a densely filled
$\Delta_{i}$ spectrum in reciprocal space. The local density $n_{i}$,
on the other hand, is quite insensitive to $U$ as it varies on the
scale of one lattice spacing, always following the lattice potential.

\begin{figure}
\includegraphics[width=0.5\columnwidth]{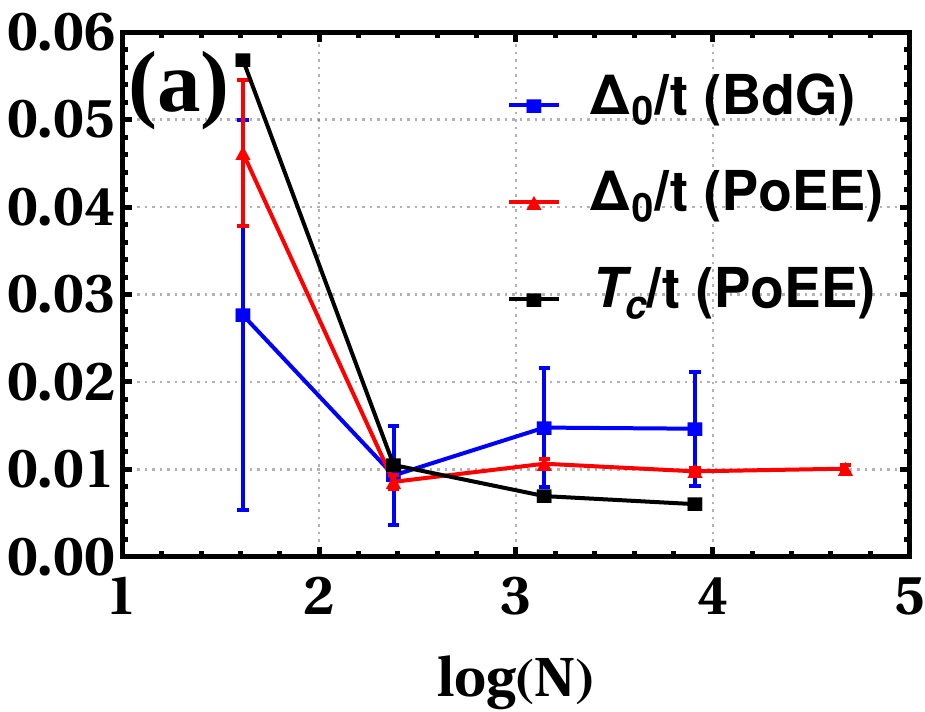}\includegraphics[width=0.5\columnwidth]{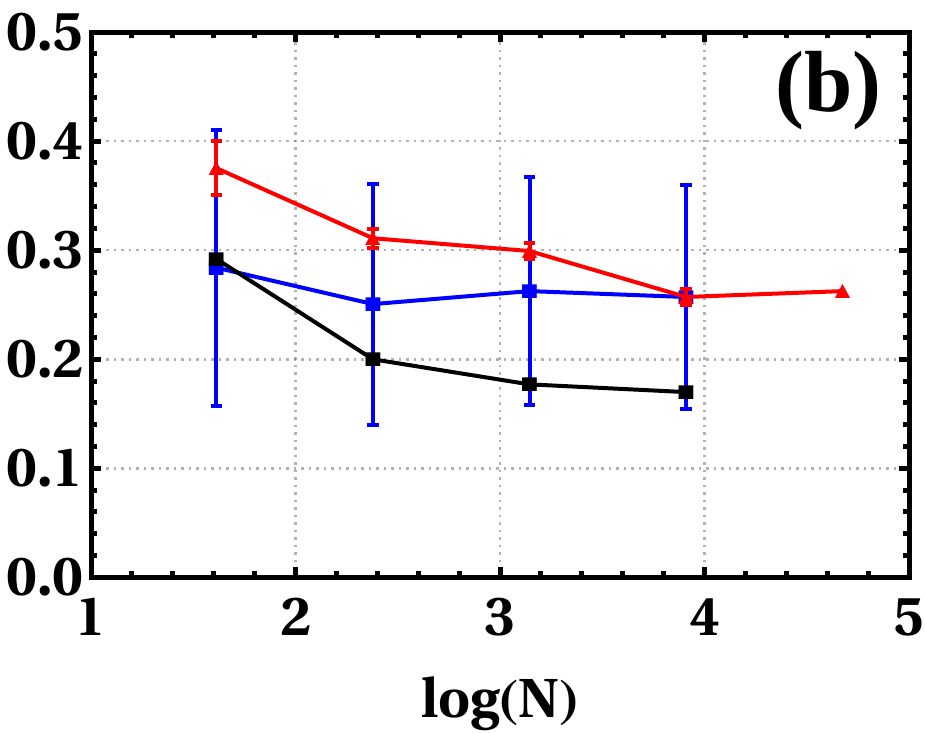}

\includegraphics[width=0.5\columnwidth]{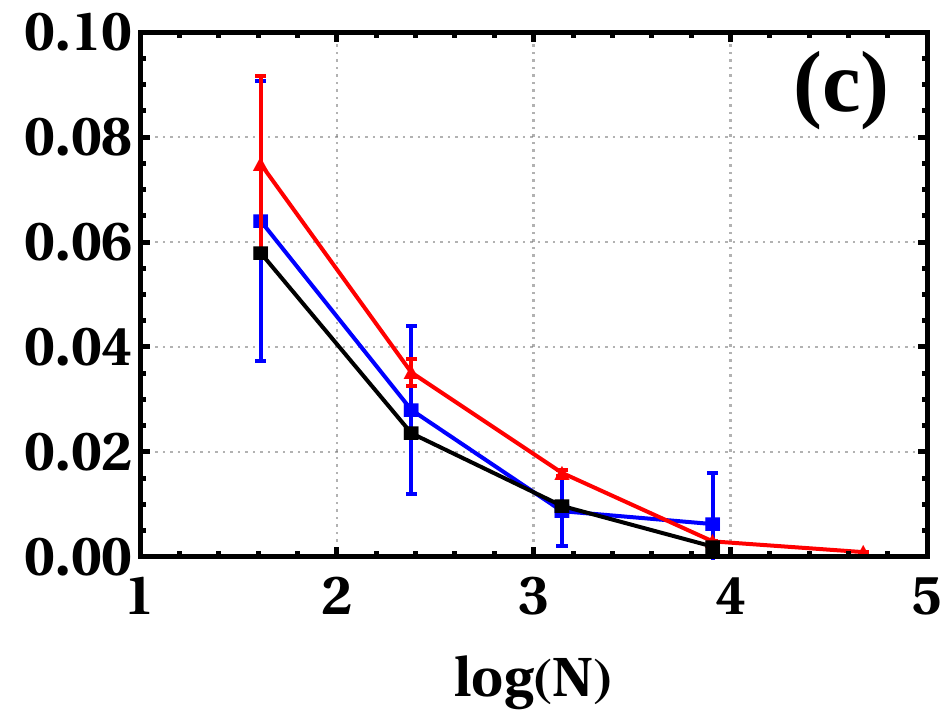}\includegraphics[width=0.5\columnwidth]{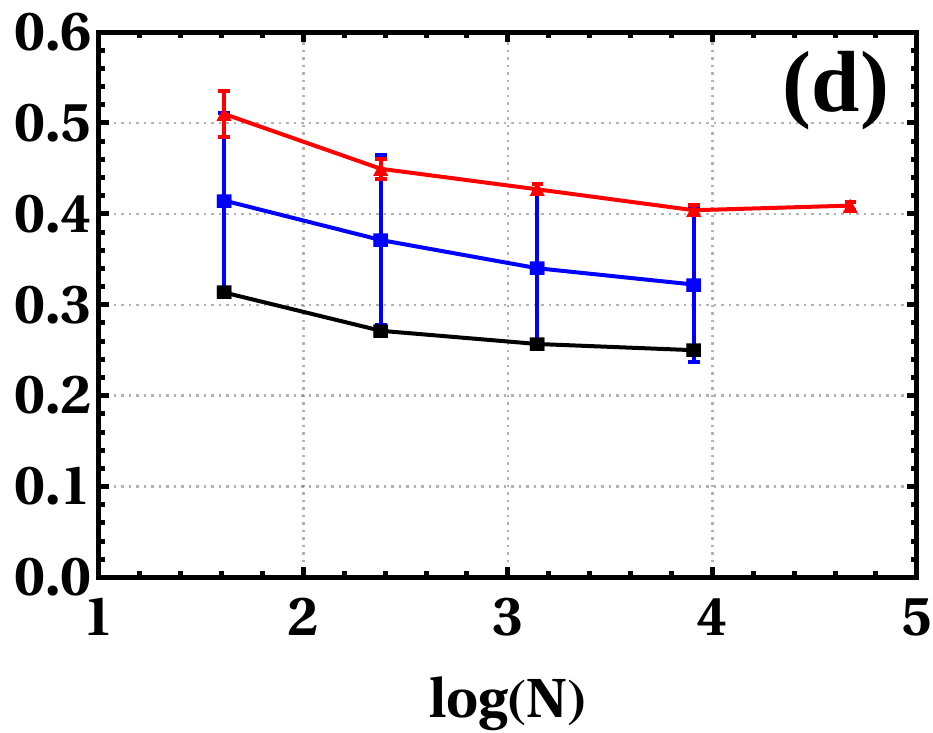}

\includegraphics[width=0.5\columnwidth]{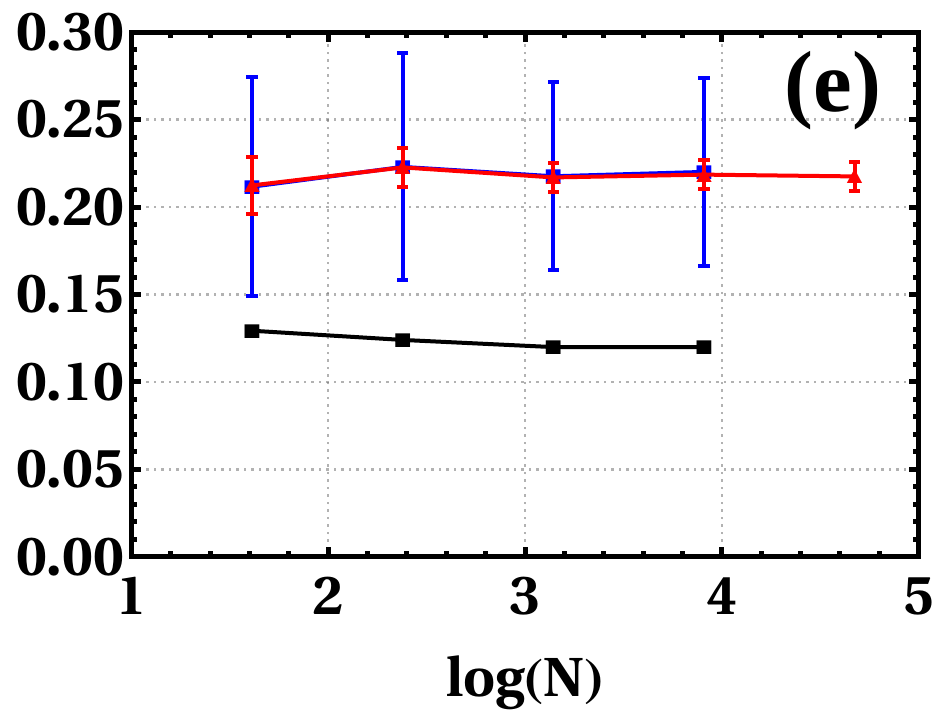}\includegraphics[width=0.5\columnwidth]{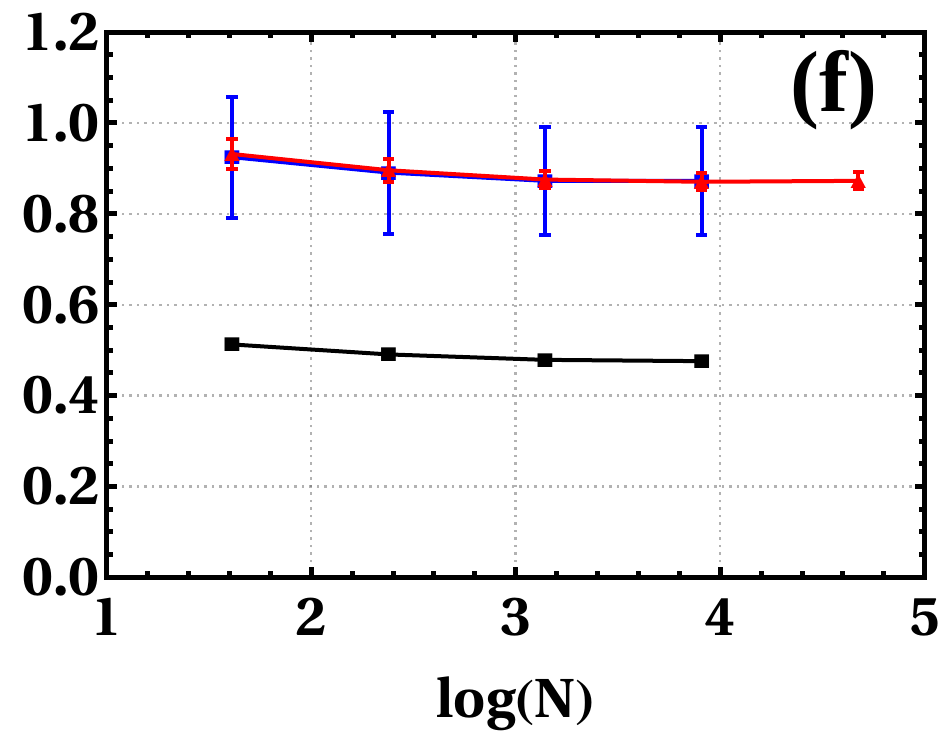}

\caption{\label{fig:scaling} Superconducting order parameter $\Delta_{0}=\sum_{i}\Delta_{i}/N$
and critical temperature $T_{c}$ as a function of the approximant
size $N$. We show the results from both the Bogoliubov-de Gennes
(BdG) $\left(\Delta_{0}\right)$ and the pairing of exact eigenstates
(PoEE) $\left(\Delta_{0},\,T_{c}\right)$ methods. (a) $U=1.5t$ and
$n=0.25$; (b) $U=3.0t$ and $n=0.25$; (c) $U=1.5t$ and $n=2/s^{2}$;
(d) $U=3.0t$ and $n=2/s^{2}$; (e) $U=1.5t$ and $n=1.00$; (f) $U=3.0t$
and $n=1.00$.}
\end{figure}

We now investigate the superconducting ground-state evolution as a
function of the approximant size $N$. On one hand---based on previous
studies considering the multifractal states observed at the Anderson
metal-insulator transition---\citep{richardella10,rodriguez10,rodriguez11}
one could naively expect an enhancement of the superconductivity\citep{feigelman07,feigelman10}
as one moves toward the infinite quasicrystal since the electronic
states become more and more critical. On the other hand, all experiments
so far find a reduction, or even a complete suppression, of $T_{c}$
as one goes from the approximant to the quasicrystal.\citep{graebner87,deguchi15,kamiya18}
In Fig. \ref{fig:scaling}, we show the superconducting order parameter
$\Delta_{0}=\sum_{i}\Delta_{i}/N$ and $T_{c}$ as a function of the
approximant size $N$. We compare the results for $\Delta_{0}$ using
the full numerical solution of the BdG equations, Eqs.\eqref{eq:self_bdg},
and that coming from the PoEE, Eqs. \eqref{eq:self_PoEE} and \eqref{eq:self_PoEE_n}
[the PoEE is cheaper numerically and allows us to go up to $N=47321$
and also to estimate $T_{c}$, see Eq. \eqref{eq:self_PoEE_linear}].
We find that $\Delta_{0}$ and $T_{c}$ remain essentially constant
for $N\ge239$ at all fillings but $n=2/s^{2}$, and the results of
both methods agree even quantitatively. This implies that the approximants
are able to capture the behavior of the infinite quasicrystal, and
that the nature of the electronic wavefunctions in the infinite
quasicrystal does not affect its superconductivity. This somewhat
disappointing result also shows that the expected analogy to disordered
systems close to the Anderson metal-insulator transition does not
hold (for all parameter sets we simulate, considering $U\ge1.5t$,
we find no enhancement superconductivity as we increase $N$). Our
scaling results for a general filling in Fig. \ref{fig:scaling} also
do not agree with the experimentally observed suppression of $T_{c}$
as $N$ increases. The only filling which captures the experimental
trend is $n=2/s^{2}$, corresponding to the pseudogap in the DOS.
Here, both $\Delta_{0}$ and $T_{c}$ are suppressed, at weak coupling,
as we increase the approximant size, due to the pseudogap in the DOS.
From Fig. \ref{fig:ipr_kohn}(b), we see that the pseudogap gets more
and more pronounced as $N$ increases, and for $N\rightarrow\infty$
we must have $U>U_{c}$, a critical coupling, for the system to display
superconductivity. For $U=1.5t$, the PoEE approach suggests that
$\Delta_{0}\rightarrow0$, whereas the full BdG solution finds $\Delta_{0}>0$,
albeit small. The difference comes from the fact that the BdG method
modifies the eigenstates of the non-interacting Hamiltonian, and it
shows that we are already above the critical coupling $U_{c}$. As
$U$ increases, all fillings behave similarly and the results are
essentially size independent. Therefore, the suppression of $T_{c}$
in a quasicrystal occurs only if the Fermi level sits at a pseudogap
--- a condition routinely met in real quasicrystals --- and at weak coupling.

\section{\label{sec:Discussion-and-connection}Discussion and connection to
experiments}

Our results show that the physics observed at the mean-field
level in randomly disordered superconductors is not present in quasicrystals.
In particular, we do not observe the formation of superconducting
islands due to the deterministic character of the lattice potential
we consider,\citep{deneau89} a conclusion which should be valid for
a broad range of quasicrystals that can be similarly constructed via
inflation or substitution rules. As was recently shown,\citep{dodaro18}
the existence of such islands circumvents Anderson's theorem\citep{anderson59,abrikosov61}
and generically enhances $T_{c}$. We also do not find an increase
of $T_{c}$ due to the multifractal nature of the electronic state in
the infinite quasicrystal, as is expected for disordered systems close
to the Anderson metal-insulator transition.\citep{feigelman07,feigelman10}
Taken together, these observations imply that the BdG solution for
superconductivity in a quasicrystal essentially fulfills Anderson's
theorem, i.e., the electrons form pairs with time-reversed eigenstates,
say $\nu\uparrow$ and $\bar{\nu}\downarrow$, of the non-interacting
model. Thus the superconductivity is of the conventional weak-coupling
BCS type, with both $T_{c}$ and $\Delta_{0}$ depending weakly on
$N$ (this point is further supported by the excellent agreement between
the results of the BdG and PoEE methods in Fig. \ref{fig:scaling}).
An interesting consequence of this observation is the fact that Anderson's
theorem also implies that $T_{c}$ does not depend on the wave functions
of $\mathcal{H}_{0}$, but only on its spectrum, and thus $T_{c}$
should be mainly governed by the DOS in Fig. \ref{fig:tiling_def}(c).
Our scaling of the order parameter $\Delta_{0}$ in Fig. \ref{fig:scaling}
illustrates this conclusion as the only distinct behavior is observed
at the pseudogap and at weak coupling. Concerning the finite temperature
critical properties of the model, the absence of rare regions, together
with the fact that Luck's criterion\citep{luck93} holds for the octagonal
tilling (as it does for most tilings constructed via inflation rules),
implies that the mean-field BCS solution is expected to hold, as is
observed experimentally.\citep{kamiya18}

Although the filling $n=2/s^{2}$ looks like a finely tuned exception
in our model, this point is actually very relevant experimentally,
since in most quasicrystals the Fermi energy is located at a pseudogap.\citep{fujiwara89,fujiwara91,ishikawa17,jazbec14}
Because of that, our results naturally account for the suppression
of superconductivity as one goes from the approximant to the quasicrystal,\citep{graebner87,deguchi15}
and for the conventional superconductivity in the Al-Zn-Mg quasicrystal,
as reported in Ref. \onlinecite{kamiya18}. One obvious implication
of our findings is that a small amount of non-magnetic impurities
should enhance $T_{c}$ in a quasicrystal, similarly to what is predicted
for superconducting semimetals.\citep{nandkishore13,potirniche14,dodaro18}
A similar possibility to increase $T_{c}$ is to dope the system,
moving the Fermi level away from the pseudogap.

We can also use our results to understand the absence of superconductivity
in $\aualyb$\citep{deguchi12} and other related heavy-fermion quasicrystals.\citep{deguchi15}
In this class of quasicrystals, a non-Fermi liquid behavior was reported
without the tuning of an external parameter.\citep{deguchi12} Interestingly,
such electronic behavior is absent in the approximants. A plausible
scenario to understand these observations is the presence of unscreened
magnetic moments down to $T\rightarrow0$ in the quasicrystals, while
in the approximants the moments are always screened below a temperature
$T^{\star}$.\citep{andrade15} Therefore, while superconductivity
is observed in the approximants of heavy-fermion quasicrystals,\citep{deguchi15}
the unscreened local moments in quasicrystals act as local pair-breaking
defects and further suppress the superconductivity, making it unlikely
for this phase to appear in an experimentally accessible temperature.

Overall, despite the fractal geometry of quasicrystals,\citep{kempkes19}
the observation of unconventional superconductivity in quasicrystals\citep{sakai17}
will probably require the same ingredients as in periodic metals,
meaning appreciable electron-electron interaction, most likely in
a material which does not involve $f$ electrons. Another interesting
problem would be to understand the compounds with a strong electron-phonon
coupling where the ubiquitous phonon spectrum of quasicrystals would
come into play,\citep{brown18} contrasting these findings with superconductivity
in elastically strained crystals where the electronic structure is
modulated in response to local lattice deformations.\citep{zhu03}

\section{Conclusions}

In this paper, we have studied the electronic properties and the $s$-wave
superconductivity in the two-dimensional Ammann-Beenker tiling. For
the electronic properties, we employed the Kohn's localization tensor
and the IPR to access the extent of the electronic
states. As in one-dimensional examples,\citep{varma16} we find that
the localization tensor gives a more detailed account on the conduction
properties of a quasiperiodic system, as shown, for instance, in
the better description of the insulating behavior expected for the
pseudogap at $n=2/s^{2}$. 

To investigate the superconductivity, we considered both real-space
BdG and the PoEE approaches
to calculate the local pairing amplitude $\Delta_{i}$. We show that
$\Delta_{i}$ is essentially determined by its local environment and
that the formation of superconducting islands is absent due to the
deterministic nature of the lattice potential we consider.\citep{deneau89}
Therefore, we find conventional BCS superconductivity in a quasicrystal,
despite the nature of their noninteracting electronic states.
The pairing mechanism is the one suggested by Anderson,\citep{anderson59}
with time-reversed eigenstates forming the Cooper pairs. In the weak-coupling
limit, the superconductivity is suppressed at the pseudogap as we
increase the approximant size. Our findings are in accordance with
recent experimental observations.\citep{kamiya18}

Since the physics of rare events, which has profound effects in random
inhomogeneous systems,\citep{thomas06} is absent in quasicrystals,
one may expect their electronic and magnetic responses to display
a more conventional behavior whenever long-range order is present.\citep{wessel03,anu04,vieira05a,thiem15,hartman16}
Of course, the local response is still highly non-trivial due to the
fractal geometry of quasicrystalline lattice,\citep{deguchi12,andrade15}
and future work on correlation effects on quasicrystals are certain
to provide many more surprises, an avenue that nowadays can also be
explored using different platforms such as cold atoms,\citep{viebahn19}
electronic systems with incommensurate order,\citep{flicker15} or
strongly correlated electronic systems at fractional filling.\citep{sagi16}
\begin{acknowledgments}
We acknowledge J. H. Garcia, J. A. Hoyos, G. F. Magno, P. B. Mendonça,
and R. T. Scalettar for useful discussions. R.N.A was supported by
the CAPES (Brazil) - Finance Code 001. E.C.A. was supported by CNPq
(Brazil) Grant No. 302065/2016-4. 
\end{acknowledgments}

\appendix

\section{\label{sec:Spectral-function}Spectral function}

To gain further insight on the electronic structure of the non-interacting
model, we calculate the spectral function 
\begin{equation}
A\left(\boldsymbol{k},\omega\right)=\sum_{\nu}\delta\left(\omega-\varepsilon_{\nu}\right)\left|\psi_{\nu}\left(\boldsymbol{k}\right)\right|^{2},\label{eq:spectral}
\end{equation}
where $\psi_{\nu}\left(\boldsymbol{k}\right)$ is the $\nu$-th eigenstate
of the non-interacting tight-binding Hamiltonian $\mathcal{H}_{0}$,
projected onto the momentum basis, and $\varepsilon_{\nu}$ is its
corresponding eigenenergy. To calculate $A\left(\boldsymbol{k},\omega\right)$
numerically, we represent the Dirac-delta function as a Lorentzian
with a broadening $\gamma=0.01t$.

\begin{figure}[t]
\includegraphics[width=1\columnwidth]{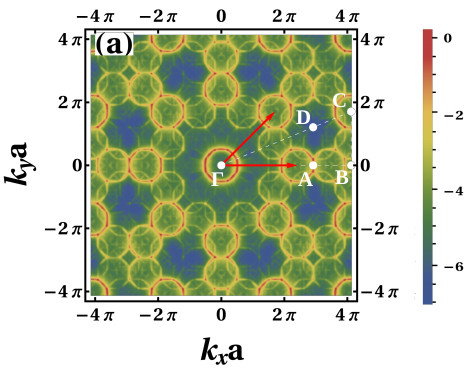}

\includegraphics[width=1\columnwidth]{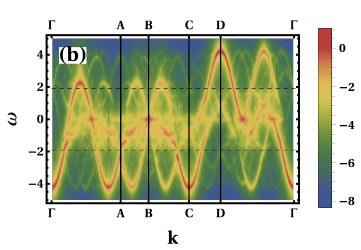}

\caption{\label{fig:arpes}(a) Surfaces of constant energy for the filling
$n=2/s^{2}$. The color intensity is determined by the spectral function
$A\left(\boldsymbol{k},\omega\right)$ and is shown in a log scale.
The vectors shown here are the ones in Fig. \ref{fig:tiling_def}(b)
connecting the brightest Bragg peaks. (b) Energy as a function of
momentum for the path defined in (a). The dashed lines set the energy
of the pseudogap. Again, the color intensity is determined by the
spectral function $A\left(\boldsymbol{k},\omega\right)$ and is shown
in a log scale. These results were obtained for the approximant with
$N=8119$ sites.}
\end{figure}

In Fig. \ref{fig:arpes}(a), we show the constant energy surfaces for
the filling $2/s^{2}$ and the eightfold rotation symmetry is evident.\citep{rotengerg00,rogalev15}
Because of the pseudogap, the Fermi surface-like contours are broken
into pockets which are centered around the brightest x-ray spots displayed
in Fig. \ref{fig:tiling_def}(b). Since we have a dense set of Bragg
peaks, there are several of these pockets and they intersect each
other, making an immediate association to a pseudogap difficult in
a finite approximant, where we superimpose our numerical broadening
to the true physical broadening of the curves, coming from the fact
that momentum is not a good quantum number (see the finite size scaling
in Fig. \ref{fig:ipr_kohn}). 

The energy as a function of momentum, for a given path in reciprocal
space, is displayed in Fig. \ref{fig:arpes}(b). There, we see that
close to the band edges the dispersionlike curves show a parabolic
behavior. Therefore, a nearly-free-electron viewpoint appears to be
a good starting point to understand the electronic properties of quasicrystals
at these extreme fillings.\citep{smith87} As we move on toward the
band center, the dispersive features become more and more blurred
due to the presence of gap openings at the crossing of the many parabolic
bandlike curves. The boundaries between these two regimes are roughly
set by the location of the pseudogap, where seemly linearly dispersing
features are present.\citep{timusk13} Precisely at the band center,
there is a flat bandlike structure, which can be directly linked
to the huge peak in the DOS in Fig. \ref{fig:tiling_def}(c). However,
one also observes dispersive features, confirming that at half-filling
the model given by Eq. \eqref{eq:h0} is metallic. 

\section{\label{sec:Pairing-of-exact}Pairing of exact eigenstates (PoEE)}

To complement the BdG results in the main text, we also consider the
so-called pairing of exact eigenstates.\citep{ghosal98,ghosal01}
This approach is a generalization of Anderson's original idea of pairing
an exact eigenstate of an inhomogeneous system to its time-reversed
pair.\citep{anderson59} Since this formalism considers only the eigenstates
and eigenenergies of Eq. \eqref{eq:h0}, it allows us to investigate
up to the $N=47321$ approximant. A full self-consistent solution
of the BdG equations for this number of sites is computationally prohibitive
within our exact diagonalization scheme, although it may be possible
using, for example, the kernel polynomial method.\citep{weisse06,covaci10}

We now briefly review the method. We start with the eigenstates $\psi_{\nu}$
of the non-interacting Hamiltonian $\mathcal{H}_{0}$ in Eq. \eqref{eq:h0}.
We then pair up electrons in time-reversed eigenstates, say $\nu\uparrow$
and $\bar{\nu}\downarrow$. The analogous BCS Hamiltonian in this
basis is then given by\citep{ma85} 
\begin{equation}
\tilde{\mathcal{H}}=\sum_{\nu,\sigma}\xi_{\nu}c_{\nu\sigma}^{\dagger}c_{\nu\sigma}-U\sum_{\nu,\zeta}M_{\nu,\zeta}c_{\nu\uparrow}^{\dagger}c_{\bar{\nu}\downarrow}^{\dagger}c_{\bar{\zeta}\downarrow}c_{\zeta\uparrow},\label{eq:h_PoEE}
\end{equation}
where the matrix $M_{\nu,\zeta}$ is given by 
\begin{equation}
M_{\nu,\zeta}=\sum_{i}\left|\psi_{\nu}\left(i\right)\right|^{2}\left|\psi_{\zeta}\left(i\right)\right|^{2},\label{eq:m_matrix}
\end{equation}
and $\xi_{\nu}=\varepsilon_{\nu}-\tilde{\mu}$ is the energy of the
non-interacting problem measured with respect to effective, Hartree-shifted,
chemical potential. A mean-field treatment of Eq. \eqref{eq:h_PoEE}
then leads to the following set of self-consistent equations:
\begin{align}
\Delta_{\nu} & =U\sum_{\zeta}M_{\nu,\zeta}\frac{\Delta_{\zeta}}{2E_{\zeta}}\tanh\text{\ensuremath{\frac{E_{\zeta}}{2T}}},\label{eq:self_PoEE}\\
\left\langle n\right\rangle  & =\frac{1}{N}\sum_{\nu}\left(1-\frac{\xi_{\nu}}{E_{\nu}}\tanh\text{\ensuremath{\frac{E_{\nu}}{2T}}}\right),\label{eq:self_PoEE_n}
\end{align}
where $E_{\nu}=\sqrt{\xi_{\nu}^{2}+\Delta_{\nu}^{2}}$ and $T$ is
the temperature. The first equation determines the pairing amplitude
of each eigenstate $\nu$, whereas the second one fixes the effective
chemical potential $\tilde{\mu}$ (notice here we are only able to
fix the average electronic density in the system). Once we solve these
equations, we can then obtain the real space pairing amplitudes at
$T=0$,\citep{ghosal01,ghosal98} and we have that $\left\langle \Delta_{i}\right\rangle =\left\langle \Delta_{\nu}\right\rangle =\Delta_{0}$.
The scaling of the order parameter is shown in Fig. \ref{fig:scaling}
and we see that the PoEE results nicely follow the full BdG solution,
thus confirming Anderson's pairing and pointing to the conventional
nature of the superconductivity in quasicrystals.

If we now linearize Eq. \eqref{eq:self_PoEE}, we get the following
set of linear equations:
\begin{align}
\Delta_{\nu} & =\sum_{\zeta}B_{\nu\zeta}\left(T\right)\Delta_{\zeta},\label{eq:self_PoEE_linear}
\end{align}
where we defined the matrix $B_{\nu,\zeta}\left(T\right)=UM_{\nu,\zeta}\tanh\left(\left|\xi_{\zeta}\right|/2T\right)/\left|\xi_{\zeta}\right|$.
$T_{c}$ is then given by temperature where the largest eigenvalue
of $B_{\nu,\zeta}\left(T\right)$ becomes equal to $1$ with $\left\langle n\right\rangle $
fixed [although $B_{\nu,\zeta}\left(T\right)$ is non-symmetric, we
checked that its eigenvalues are real]. The resulting $T_{c}$ are
also shown in Fig. \ref{fig:scaling}, and we can compare it with
the BdG results for $N=1393$ in Figs. \ref{fig:delta_i} (c) and \ref{fig:delta_i} (d).
For $n=0.25$, we have $T_{c}/t=0.011$ (BdG) and $0.008$ (PoEE)
whereas for $n=1.00$ we have $T_{c}/t=0.137$ (BdG) and $0.120$
(PoEE), again showing a good agreement between the methods.


\begin{thebibliography}{83}%
\makeatletter
\providecommand \@ifxundefined [1]{%
 \@ifx{#1\undefined}
}%
\providecommand \@ifnum [1]{%
 \ifnum #1\expandafter \@firstoftwo
 \else \expandafter \@secondoftwo
 \fi
}%
\providecommand \@ifx [1]{%
 \ifx #1\expandafter \@firstoftwo
 \else \expandafter \@secondoftwo
 \fi
}%
\providecommand \natexlab [1]{#1}%
\providecommand \enquote  [1]{``#1''}%
\providecommand \bibnamefont  [1]{#1}%
\providecommand \bibfnamefont [1]{#1}%
\providecommand \citenamefont [1]{#1}%
\providecommand \href@noop [0]{\@secondoftwo}%
\providecommand \href [0]{\begingroup \@sanitize@url \@href}%
\providecommand \@href[1]{\@@startlink{#1}\@@href}%
\providecommand \@@href[1]{\endgroup#1\@@endlink}%
\providecommand \@sanitize@url [0]{\catcode `\\12\catcode `\$12\catcode
  `\&12\catcode `\#12\catcode `\^12\catcode `\_12\catcode `\%12\relax}%
\providecommand \@@startlink[1]{}%
\providecommand \@@endlink[0]{}%
\providecommand \url  [0]{\begingroup\@sanitize@url \@url }%
\providecommand \@url [1]{\endgroup\@href {#1}{\urlprefix }}%
\providecommand \urlprefix  [0]{URL }%
\providecommand \Eprint [0]{\href }%
\providecommand \doibase [0]{http://dx.doi.org/}%
\providecommand \selectlanguage [0]{\@gobble}%
\providecommand \bibinfo  [0]{\@secondoftwo}%
\providecommand \bibfield  [0]{\@secondoftwo}%
\providecommand \translation [1]{[#1]}%
\providecommand \BibitemOpen [0]{}%
\providecommand \bibitemStop [0]{}%
\providecommand \bibitemNoStop [0]{.\EOS\space}%
\providecommand \EOS [0]{\spacefactor3000\relax}%
\providecommand \BibitemShut  [1]{\csname bibitem#1\endcsname}%
\let\auto@bib@innerbib\@empty
\bibitem [{\citenamefont {Shechtman}\ \emph {et~al.}(1984)\citenamefont
  {Shechtman}, \citenamefont {Blech}, \citenamefont {Gratias},\ and\
  \citenamefont {Cahn}}]{shechtman84}%
  \BibitemOpen
  \bibfield  {author} {\bibinfo {author} {\bibfnamefont {D.}~\bibnamefont
  {Shechtman}}, \bibinfo {author} {\bibfnamefont {I.}~\bibnamefont {Blech}},
  \bibinfo {author} {\bibfnamefont {D.}~\bibnamefont {Gratias}}, \ and\
  \bibinfo {author} {\bibfnamefont {J.~W.}\ \bibnamefont {Cahn}},\ }\bibfield
  {title} {\enquote {\bibinfo {title} {{Metallic Phase with Long-Range
  Orientational Order and No Translational Symmetry}},}\ }\href {\doibase
  10.1103/PhysRevLett.53.1951} {\bibfield  {journal} {\bibinfo  {journal}
  {Phys. Rev. Lett.}\ }\textbf {\bibinfo {volume} {53}},\ \bibinfo {pages}
  {1951} (\bibinfo {year} {1984})}\BibitemShut {NoStop}%
\bibitem [{\citenamefont {Levine}\ and\ \citenamefont
  {Steinhardt}(1984)}]{levine84}%
  \BibitemOpen
  \bibfield  {author} {\bibinfo {author} {\bibfnamefont {D.}~\bibnamefont
  {Levine}}\ and\ \bibinfo {author} {\bibfnamefont {P.~J.}\ \bibnamefont
  {Steinhardt}},\ }\bibfield  {title} {\enquote {\bibinfo {title}
  {{Quasicrystals: A New Class of Ordered Structures}},}\ }\href {\doibase
  10.1103/PhysRevLett.53.2477} {\bibfield  {journal} {\bibinfo  {journal}
  {Phys. Rev. Lett.}\ }\textbf {\bibinfo {volume} {53}},\ \bibinfo {pages}
  {2477} (\bibinfo {year} {1984})}\BibitemShut {NoStop}%
\bibitem [{\citenamefont {Grimm}\ and\ \citenamefont
  {Schreiber}(2003)}]{grimm03}%
  \BibitemOpen
  \bibfield  {author} {\bibinfo {author} {\bibfnamefont {U.}~\bibnamefont
  {Grimm}}\ and\ \bibinfo {author} {\bibfnamefont {M.}~\bibnamefont
  {Schreiber}},\ }\bibfield  {title} {\enquote {\bibinfo {title} {{Energy
  spectra and eigenstates of quasiperiodic tight-binding Hamiltonians}},}\ }in\
  \href@noop {} {\emph {\bibinfo {booktitle} {{Quasicrystals - Structure and
  Physical Properties}}}},\ \bibinfo {editor} {edited by\ \bibinfo {editor}
  {\bibfnamefont {H.-R.}\ \bibnamefont {Trebin}}}\ (\bibinfo  {publisher}
  {Wiley-VCH, Weinheim},\ \bibinfo {year} {2003})\BibitemShut {NoStop}%
\bibitem [{\citenamefont {Jagannathan}\ and\ \citenamefont
  {Pi\'{e}chon}(2007)}]{anu06}%
  \BibitemOpen
  \bibfield  {author} {\bibinfo {author} {\bibfnamefont {A.}~\bibnamefont
  {Jagannathan}}\ and\ \bibinfo {author} {\bibfnamefont {F.}~\bibnamefont
  {Pi\'{e}chon}},\ }\bibfield  {title} {\enquote {\bibinfo {title} {{Energy
  levels and their correlations in quasicrystals}},}\ }\href {\doibase
  10.1080/14786430701196990} {\bibfield  {journal} {\bibinfo  {journal}
  {Philos. Mag.}\ }\textbf {\bibinfo {volume} {87}},\ \bibinfo {pages} {2389}
  (\bibinfo {year} {2007})}\BibitemShut {NoStop}%
\bibitem [{\citenamefont {Kohmoto}\ \emph {et~al.}(1987)\citenamefont
  {Kohmoto}, \citenamefont {Sutherland},\ and\ \citenamefont
  {Tang}}]{kohmoto87}%
  \BibitemOpen
  \bibfield  {author} {\bibinfo {author} {\bibfnamefont {M.}~\bibnamefont
  {Kohmoto}}, \bibinfo {author} {\bibfnamefont {B.}~\bibnamefont {Sutherland}},
  \ and\ \bibinfo {author} {\bibfnamefont {C.}~\bibnamefont {Tang}},\
  }\bibfield  {title} {\enquote {\bibinfo {title} {{Critical wave functions and
  a Cantor-set spectrum of a one-dimensional quasicrystal model}},}\ }\href
  {\doibase 10.1103/PhysRevB.35.1020} {\bibfield  {journal} {\bibinfo
  {journal} {Phys. Rev. B}\ }\textbf {\bibinfo {volume} {35}},\ \bibinfo
  {pages} {1020} (\bibinfo {year} {1987})}\BibitemShut {NoStop}%
\bibitem [{\citenamefont {Tsunetsugu}\ \emph {et~al.}(1991)\citenamefont
  {Tsunetsugu}, \citenamefont {Fujiwara}, \citenamefont {Ueda},\ and\
  \citenamefont {Tokihiro}}]{tsunetsugu91a}%
  \BibitemOpen
  \bibfield  {author} {\bibinfo {author} {\bibfnamefont {H.}~\bibnamefont
  {Tsunetsugu}}, \bibinfo {author} {\bibfnamefont {T.}~\bibnamefont
  {Fujiwara}}, \bibinfo {author} {\bibfnamefont {K.}~\bibnamefont {Ueda}}, \
  and\ \bibinfo {author} {\bibfnamefont {T.}~\bibnamefont {Tokihiro}},\
  }\bibfield  {title} {\enquote {\bibinfo {title} {{Electronic properties of
  the Penrose lattice. I. Energy spectrum and wave functions}},}\ }\href
  {\doibase 10.1103/PhysRevB.43.8879} {\bibfield  {journal} {\bibinfo
  {journal} {Phys. Rev. B}\ }\textbf {\bibinfo {volume} {43}},\ \bibinfo
  {pages} {8879} (\bibinfo {year} {1991})}\BibitemShut {NoStop}%
\bibitem [{\citenamefont {Yuan}\ \emph {et~al.}(2000)\citenamefont {Yuan},
  \citenamefont {Grimm}, \citenamefont {Repetowicz},\ and\ \citenamefont
  {Schreiber}}]{yuan00}%
  \BibitemOpen
  \bibfield  {author} {\bibinfo {author} {\bibfnamefont {H.~Q.}\ \bibnamefont
  {Yuan}}, \bibinfo {author} {\bibfnamefont {U.}~\bibnamefont {Grimm}},
  \bibinfo {author} {\bibfnamefont {P.}~\bibnamefont {Repetowicz}}, \ and\
  \bibinfo {author} {\bibfnamefont {M.}~\bibnamefont {Schreiber}},\ }\bibfield
  {title} {\enquote {\bibinfo {title} {{Energy spectra, wave functions, and
  quantum diffusion for quasiperiodic systems}},}\ }\href {\doibase
  10.1103/PhysRevB.62.15569} {\bibfield  {journal} {\bibinfo  {journal} {Phys.
  Rev. B}\ }\textbf {\bibinfo {volume} {62}},\ \bibinfo {pages} {15569}
  (\bibinfo {year} {2000})}\BibitemShut {NoStop}%
\bibitem [{\citenamefont {Tanese}\ \emph {et~al.}(2014)\citenamefont {Tanese},
  \citenamefont {Gurevich}, \citenamefont {Baboux}, \citenamefont {Jacqmin},
  \citenamefont {Lema\^{i}tre}, \citenamefont {Galopin}, \citenamefont
  {Sagnes}, \citenamefont {Amo}, \citenamefont {Bloch},\ and\ \citenamefont
  {Akkermans}}]{tanese14}%
  \BibitemOpen
  \bibfield  {author} {\bibinfo {author} {\bibfnamefont {D.}~\bibnamefont
  {Tanese}}, \bibinfo {author} {\bibfnamefont {E.}~\bibnamefont {Gurevich}},
  \bibinfo {author} {\bibfnamefont {F.}~\bibnamefont {Baboux}}, \bibinfo
  {author} {\bibfnamefont {T.}~\bibnamefont {Jacqmin}}, \bibinfo {author}
  {\bibfnamefont {A.}~\bibnamefont {Lema\^{i}tre}}, \bibinfo {author}
  {\bibfnamefont {E.}~\bibnamefont {Galopin}}, \bibinfo {author} {\bibfnamefont
  {I.}~\bibnamefont {Sagnes}}, \bibinfo {author} {\bibfnamefont
  {A.}~\bibnamefont {Amo}}, \bibinfo {author} {\bibfnamefont {J.}~\bibnamefont
  {Bloch}}, \ and\ \bibinfo {author} {\bibfnamefont {E.}~\bibnamefont
  {Akkermans}},\ }\bibfield  {title} {\enquote {\bibinfo {title} {{Fractal
  Energy Spectrum of a Polariton Gas in a Fibonacci Quasiperiodic
  Potential}},}\ }\href {\doibase 10.1103/PhysRevLett.112.146404} {\bibfield
  {journal} {\bibinfo  {journal} {Phys. Rev. Lett.}\ }\textbf {\bibinfo
  {volume} {112}},\ \bibinfo {pages} {146404} (\bibinfo {year}
  {2014})}\BibitemShut {NoStop}%
\bibitem [{\citenamefont {Mac\'{e}}\ \emph {et~al.}(2017)\citenamefont
  {Mac\'{e}}, \citenamefont {Jagannathan}, \citenamefont {Kalugin},
  \citenamefont {Mosseri},\ and\ \citenamefont {Pi\'{e}chon}}]{mace17}%
  \BibitemOpen
  \bibfield  {author} {\bibinfo {author} {\bibfnamefont {N.}~\bibnamefont
  {Mac\'{e}}}, \bibinfo {author} {\bibfnamefont {A.}~\bibnamefont
  {Jagannathan}}, \bibinfo {author} {\bibfnamefont {P.}~\bibnamefont
  {Kalugin}}, \bibinfo {author} {\bibfnamefont {R.}~\bibnamefont {Mosseri}}, \
  and\ \bibinfo {author} {\bibfnamefont {F.}~\bibnamefont {Pi\'{e}chon}},\
  }\bibfield  {title} {\enquote {\bibinfo {title} {{Critical eigenstates and
  their properties in one- and two-dimensional quasicrystals}},}\ }\href
  {\doibase 10.1103/PhysRevB.96.045138} {\bibfield  {journal} {\bibinfo
  {journal} {Phys. Rev. B}\ }\textbf {\bibinfo {volume} {96}},\ \bibinfo
  {pages} {045138} (\bibinfo {year} {2017})}\BibitemShut {NoStop}%
\bibitem [{\citenamefont {Kohmoto}\ and\ \citenamefont
  {Sutherland}(1986)}]{kohmoto86}%
  \BibitemOpen
  \bibfield  {author} {\bibinfo {author} {\bibfnamefont {M.}~\bibnamefont
  {Kohmoto}}\ and\ \bibinfo {author} {\bibfnamefont {B.}~\bibnamefont
  {Sutherland}},\ }\bibfield  {title} {\enquote {\bibinfo {title} {{Electronic
  States on a Penrose Lattice}},}\ }\href {\doibase
  10.1103/PhysRevLett.56.2740} {\bibfield  {journal} {\bibinfo  {journal}
  {Phys. Rev. Lett.}\ }\textbf {\bibinfo {volume} {56}},\ \bibinfo {pages}
  {2740} (\bibinfo {year} {1986})}\BibitemShut {NoStop}%
\bibitem [{\citenamefont {Arai}\ \emph {et~al.}(1988)\citenamefont {Arai},
  \citenamefont {Tokihiro}, \citenamefont {Fujiwara},\ and\ \citenamefont
  {Kohmoto}}]{arai88}%
  \BibitemOpen
  \bibfield  {author} {\bibinfo {author} {\bibfnamefont {M.}~\bibnamefont
  {Arai}}, \bibinfo {author} {\bibfnamefont {T.}~\bibnamefont {Tokihiro}},
  \bibinfo {author} {\bibfnamefont {T.}~\bibnamefont {Fujiwara}}, \ and\
  \bibinfo {author} {\bibfnamefont {M.}~\bibnamefont {Kohmoto}},\ }\bibfield
  {title} {\enquote {\bibinfo {title} {{Strictly localized states on a
  two-dimensional Penrose lattice}},}\ }\href {\doibase
  10.1103/PhysRevB.38.1621} {\bibfield  {journal} {\bibinfo  {journal} {Phys.
  Rev. B}\ }\textbf {\bibinfo {volume} {38}},\ \bibinfo {pages} {1621}
  (\bibinfo {year} {1988})}\BibitemShut {NoStop}%
\bibitem [{\citenamefont {Rieth}\ and\ \citenamefont
  {Schreiber}(1995)}]{rieth95}%
  \BibitemOpen
  \bibfield  {author} {\bibinfo {author} {\bibfnamefont {T.}~\bibnamefont
  {Rieth}}\ and\ \bibinfo {author} {\bibfnamefont {M.}~\bibnamefont
  {Schreiber}},\ }\bibfield  {title} {\enquote {\bibinfo {title}
  {{Identification of spatially confined states in two-dimensional
  quasiperiodic lattices}},}\ }\href {\doibase 10.1103/PhysRevB.51.15827}
  {\bibfield  {journal} {\bibinfo  {journal} {Phys. Rev. B}\ }\textbf {\bibinfo
  {volume} {51}},\ \bibinfo {pages} {15827} (\bibinfo {year}
  {1995})}\BibitemShut {NoStop}%
\bibitem [{\citenamefont {Fujiwara}(1989)}]{fujiwara89}%
  \BibitemOpen
  \bibfield  {author} {\bibinfo {author} {\bibfnamefont {T.}~\bibnamefont
  {Fujiwara}},\ }\bibfield  {title} {\enquote {\bibinfo {title} {{Electronic
  structure in the Al-Mn alloy crystalline analog of quasicrystals}},}\ }\href
  {\doibase 10.1103/PhysRevB.40.942} {\bibfield  {journal} {\bibinfo  {journal}
  {Phys. Rev. B}\ }\textbf {\bibinfo {volume} {40}},\ \bibinfo {pages} {942}
  (\bibinfo {year} {1989})}\BibitemShut {NoStop}%
\bibitem [{\citenamefont {Fujiwara}\ and\ \citenamefont
  {Yokokawa}(1991)}]{fujiwara91}%
  \BibitemOpen
  \bibfield  {author} {\bibinfo {author} {\bibfnamefont {T.}~\bibnamefont
  {Fujiwara}}\ and\ \bibinfo {author} {\bibfnamefont {T.}~\bibnamefont
  {Yokokawa}},\ }\bibfield  {title} {\enquote {\bibinfo {title} {{Universal
  pseudogap at Fermi energy in quasicrystals}},}\ }\href {\doibase
  10.1103/PhysRevLett.66.333} {\bibfield  {journal} {\bibinfo  {journal} {Phys.
  Rev. Lett.}\ }\textbf {\bibinfo {volume} {66}},\ \bibinfo {pages} {333}
  (\bibinfo {year} {1991})}\BibitemShut {NoStop}%
\bibitem [{\citenamefont {Ishikawa}\ \emph {et~al.}(2017)\citenamefont
  {Ishikawa}, \citenamefont {Takagiwa}, \citenamefont {Kimura},\ and\
  \citenamefont {Tamura}}]{ishikawa17}%
  \BibitemOpen
  \bibfield  {author} {\bibinfo {author} {\bibfnamefont {A.}~\bibnamefont
  {Ishikawa}}, \bibinfo {author} {\bibfnamefont {Y.}~\bibnamefont {Takagiwa}},
  \bibinfo {author} {\bibfnamefont {K.}~\bibnamefont {Kimura}}, \ and\ \bibinfo
  {author} {\bibfnamefont {R.}~\bibnamefont {Tamura}},\ }\bibfield  {title}
  {\enquote {\bibinfo {title} {{Probing of the pseudogap via thermoelectric
  properties in the Au-Al-Gd quasicrystal approximant}},}\ }\href {\doibase
  10.1103/PhysRevB.95.104201} {\bibfield  {journal} {\bibinfo  {journal} {Phys.
  Rev. B}\ }\textbf {\bibinfo {volume} {95}},\ \bibinfo {pages} {104201}
  (\bibinfo {year} {2017})}\BibitemShut {NoStop}%
\bibitem [{\citenamefont {Jazbec}\ \emph {et~al.}(2014)\citenamefont {Jazbec},
  \citenamefont {Vrtnik}, \citenamefont {Jagli\u{c}i\'{c}}, \citenamefont
  {Kashimoto}, \citenamefont {Ivkov}, \citenamefont {Pop\u{c}evi\'{c}},
  \citenamefont {Smontara}, \citenamefont {Kim}, \citenamefont {Kim},\ and\
  \citenamefont {Dolin\u{s}ek}}]{jazbec14}%
  \BibitemOpen
  \bibfield  {author} {\bibinfo {author} {\bibfnamefont {S.}~\bibnamefont
  {Jazbec}}, \bibinfo {author} {\bibfnamefont {S.}~\bibnamefont {Vrtnik}},
  \bibinfo {author} {\bibfnamefont {Z.}~\bibnamefont {Jagli\u{c}i\'{c}}},
  \bibinfo {author} {\bibfnamefont {S.}~\bibnamefont {Kashimoto}}, \bibinfo
  {author} {\bibfnamefont {J.}~\bibnamefont {Ivkov}}, \bibinfo {author}
  {\bibfnamefont {P.}~\bibnamefont {Pop\u{c}evi\'{c}}}, \bibinfo {author}
  {\bibfnamefont {A.}~\bibnamefont {Smontara}}, \bibinfo {author}
  {\bibfnamefont {Hae~Jin}\ \bibnamefont {Kim}}, \bibinfo {author}
  {\bibfnamefont {Jin~Gyu}\ \bibnamefont {Kim}}, \ and\ \bibinfo {author}
  {\bibfnamefont {J.}~\bibnamefont {Dolin\u{s}ek}},\ }\bibfield  {title}
  {\enquote {\bibinfo {title} {{Electronic density of states and metastability
  of icosahedral Au-Al-Yb quasicrystal}},}\ }\href {\doibase
  10.1016/j.jallcom.2013.10.073} {\bibfield  {journal} {\bibinfo  {journal} {J.
  Alloys Compd.}\ }\textbf {\bibinfo {volume} {586}},\ \bibinfo {pages} {343}
  (\bibinfo {year} {2014})}\BibitemShut {NoStop}%
\bibitem [{\citenamefont {Pierce}\ \emph {et~al.}(1993)\citenamefont {Pierce},
  \citenamefont {Poon},\ and\ \citenamefont {Guo}}]{pierce93}%
  \BibitemOpen
  \bibfield  {author} {\bibinfo {author} {\bibfnamefont {F.~S.}\ \bibnamefont
  {Pierce}}, \bibinfo {author} {\bibfnamefont {S.~J.}\ \bibnamefont {Poon}}, \
  and\ \bibinfo {author} {\bibfnamefont {Q.}~\bibnamefont {Guo}},\ }\bibfield
  {title} {\enquote {\bibinfo {title} {{Electron Localization in Metallic
  Quasicrystals}},}\ }\href {\doibase 10.1126/science.261.5122.737} {\bibfield
  {journal} {\bibinfo  {journal} {Science}\ }\textbf {\bibinfo {volume}
  {261}},\ \bibinfo {pages} {737} (\bibinfo {year} {1993})}\BibitemShut
  {NoStop}%
\bibitem [{\citenamefont {{Trambly de Laissardi\`{e}re}}\ and\ \citenamefont
  {Fujiwara}(1994)}]{trambly94}%
  \BibitemOpen
  \bibfield  {author} {\bibinfo {author} {\bibfnamefont {G.}~\bibnamefont
  {{Trambly de Laissardi\`{e}re}}}\ and\ \bibinfo {author} {\bibfnamefont
  {T.}~\bibnamefont {Fujiwara}},\ }\bibfield  {title} {\enquote {\bibinfo
  {title} {{Electronic structure and transport in a model approximant of the
  decagonal quasicrystal Al-Cu-Co}},}\ }\href {\doibase
  10.1103/PhysRevB.50.9843} {\bibfield  {journal} {\bibinfo  {journal} {Phys.
  Rev. B}\ }\textbf {\bibinfo {volume} {50}},\ \bibinfo {pages} {9843}
  (\bibinfo {year} {1994})}\BibitemShut {NoStop}%
\bibitem [{\citenamefont {Timusk}\ \emph {et~al.}(2013)\citenamefont {Timusk},
  \citenamefont {Carbotte}, \citenamefont {Homes}, \citenamefont {Basov},\ and\
  \citenamefont {Sharapov}}]{timusk13}%
  \BibitemOpen
  \bibfield  {author} {\bibinfo {author} {\bibfnamefont {T.}~\bibnamefont
  {Timusk}}, \bibinfo {author} {\bibfnamefont {J.~P.}\ \bibnamefont
  {Carbotte}}, \bibinfo {author} {\bibfnamefont {C.~C.}\ \bibnamefont {Homes}},
  \bibinfo {author} {\bibfnamefont {D.~N.}\ \bibnamefont {Basov}}, \ and\
  \bibinfo {author} {\bibfnamefont {S.~G.}\ \bibnamefont {Sharapov}},\
  }\bibfield  {title} {\enquote {\bibinfo {title} {{Three-dimensional Dirac
  fermions in quasicrystals as seen via optical conductivity}},}\ }\href
  {\doibase 10.1103/PhysRevB.87.235121} {\bibfield  {journal} {\bibinfo
  {journal} {Phys. Rev. B}\ }\textbf {\bibinfo {volume} {87}},\ \bibinfo
  {pages} {235121} (\bibinfo {year} {2013})}\BibitemShut {NoStop}%
\bibitem [{\citenamefont {de~Laissardi\`{e}re}\ and\ \citenamefont
  {Mayou}(2014)}]{deLa14}%
  \BibitemOpen
  \bibfield  {author} {\bibinfo {author} {\bibfnamefont {G.~T.}\ \bibnamefont
  {de~Laissardi\`{e}re}}\ and\ \bibinfo {author} {\bibfnamefont
  {D.}~\bibnamefont {Mayou}},\ }\bibfield  {title} {\enquote {\bibinfo {title}
  {{Anomalous electronic transport in quasicrystals and related complex
  metallic alloys}},}\ }\href {\doibase 10.1016/j.crhy.2013.09.010} {\bibfield
  {journal} {\bibinfo  {journal} {C. R. Physique}\ }\textbf {\bibinfo {volume}
  {15}},\ \bibinfo {pages} {70} (\bibinfo {year} {2014})}\BibitemShut {NoStop}%
\bibitem [{\citenamefont {Wessel}\ \emph {et~al.}(2003)\citenamefont {Wessel},
  \citenamefont {Jagannathan},\ and\ \citenamefont {Haas}}]{wessel03}%
  \BibitemOpen
  \bibfield  {author} {\bibinfo {author} {\bibfnamefont {S.}~\bibnamefont
  {Wessel}}, \bibinfo {author} {\bibfnamefont {A.}~\bibnamefont {Jagannathan}},
  \ and\ \bibinfo {author} {\bibfnamefont {S.}~\bibnamefont {Haas}},\
  }\bibfield  {title} {\enquote {\bibinfo {title} {{Quantum Antiferromagnetism
  in Quasicrystals}},}\ }\href {\doibase 10.1103/PhysRevLett.90.177205}
  {\bibfield  {journal} {\bibinfo  {journal} {Phys. Rev. Lett.}\ }\textbf
  {\bibinfo {volume} {90}},\ \bibinfo {pages} {177205} (\bibinfo {year}
  {2003})}\BibitemShut {NoStop}%
\bibitem [{\citenamefont {Vieira}(2005)}]{vieira05a}%
  \BibitemOpen
  \bibfield  {author} {\bibinfo {author} {\bibfnamefont {A.~P.}\ \bibnamefont
  {Vieira}},\ }\bibfield  {title} {\enquote {\bibinfo {title} {{Low-Energy
  Properties of Aperiodic Quantum Spin Chains}},}\ }\href {\doibase
  10.1103/PhysRevLett.94.077201} {\bibfield  {journal} {\bibinfo  {journal}
  {Phys. Rev. Lett.}\ }\textbf {\bibinfo {volume} {94}},\ \bibinfo {pages}
  {077201} (\bibinfo {year} {2005})}\BibitemShut {NoStop}%
\bibitem [{\citenamefont {Thiem}\ and\ \citenamefont
  {Chalker}(2015)}]{thiem15}%
  \BibitemOpen
  \bibfield  {author} {\bibinfo {author} {\bibfnamefont {S.}~\bibnamefont
  {Thiem}}\ and\ \bibinfo {author} {\bibfnamefont {J.~T.}\ \bibnamefont
  {Chalker}},\ }\bibfield  {title} {\enquote {\bibinfo {title} {{Long-range
  magnetic order in models for rare-earth quasicrystals}},}\ }\href {\doibase
  10.1103/PhysRevB.92.224409} {\bibfield  {journal} {\bibinfo  {journal} {Phys.
  Rev. B}\ }\textbf {\bibinfo {volume} {92}},\ \bibinfo {pages} {224409}
  (\bibinfo {year} {2015})}\BibitemShut {NoStop}%
\bibitem [{\citenamefont {Hartman}\ \emph {et~al.}(2016)\citenamefont
  {Hartman}, \citenamefont {Chiu},\ and\ \citenamefont
  {Scalettar}}]{hartman16}%
  \BibitemOpen
  \bibfield  {author} {\bibinfo {author} {\bibfnamefont {N.}~\bibnamefont
  {Hartman}}, \bibinfo {author} {\bibfnamefont {W.-T.}\ \bibnamefont {Chiu}}, \
  and\ \bibinfo {author} {\bibfnamefont {R.~T.}\ \bibnamefont {Scalettar}},\
  }\bibfield  {title} {\enquote {\bibinfo {title} {{Magnetic correlations in a
  periodic Anderson model with nonuniform conduction electron coordination}},}\
  }\href {\doibase 10.1103/PhysRevB.93.235143} {\bibfield  {journal} {\bibinfo
  {journal} {Phys. Rev. B}\ }\textbf {\bibinfo {volume} {93}},\ \bibinfo
  {pages} {235143} (\bibinfo {year} {2016})}\BibitemShut {NoStop}%
\bibitem [{\citenamefont {Koga}\ and\ \citenamefont
  {Tsunetsugu}(2017)}]{koga17}%
  \BibitemOpen
  \bibfield  {author} {\bibinfo {author} {\bibfnamefont {A.}~\bibnamefont
  {Koga}}\ and\ \bibinfo {author} {\bibfnamefont {H.}~\bibnamefont
  {Tsunetsugu}},\ }\bibfield  {title} {\enquote {\bibinfo {title}
  {{Antiferromagnetic order in the Hubbard model on the Penrose lattice}},}\
  }\href {\doibase 10.1103/PhysRevB.96.214402} {\bibfield  {journal} {\bibinfo
  {journal} {Phys. Rev. B}\ }\textbf {\bibinfo {volume} {96}},\ \bibinfo
  {pages} {214402} (\bibinfo {year} {2017})}\BibitemShut {NoStop}%
\bibitem [{\citenamefont {Luck}(1993)}]{luck93}%
  \BibitemOpen
  \bibfield  {author} {\bibinfo {author} {\bibfnamefont {J.~M.}\ \bibnamefont
  {Luck}},\ }\bibfield  {title} {\enquote {\bibinfo {title} {{A Classification
  of Critical Phenomena on Quasi-Crystals and Other Aperiodic Structures}},}\
  }\href {\doibase 10.1209/0295-5075/24/5/007} {\bibfield  {journal} {\bibinfo
  {journal} {Europhys. Lett.}\ }\textbf {\bibinfo {volume} {24}},\ \bibinfo
  {pages} {359} (\bibinfo {year} {1993})}\BibitemShut {NoStop}%
\bibitem [{\citenamefont {{K. Deguchi}}\ \emph {et~al.}(2012)\citenamefont {{K.
  Deguchi}}, \citenamefont {{S. Matsukawa}}, \citenamefont {{N. K. Sato}},
  \citenamefont {{T. Hattori}}, \citenamefont {{K. Ishida}}, \citenamefont {{H.
  Takakura}},\ and\ \citenamefont {{T. Ishimasa}}}]{deguchi12}%
  \BibitemOpen
  \bibfield  {author} {\bibinfo {author} {\bibnamefont {{K. Deguchi}}},
  \bibinfo {author} {\bibnamefont {{S. Matsukawa}}}, \bibinfo {author}
  {\bibnamefont {{N. K. Sato}}}, \bibinfo {author} {\bibnamefont {{T.
  Hattori}}}, \bibinfo {author} {\bibnamefont {{K. Ishida}}}, \bibinfo {author}
  {\bibnamefont {{H. Takakura}}}, \ and\ \bibinfo {author} {\bibnamefont {{T.
  Ishimasa}}},\ }\bibfield  {title} {\enquote {\bibinfo {title} {{Quantum
  critical state in a magnetic quasicrystal}},}\ }\href {\doibase
  10.1038/nmat3432} {\bibfield  {journal} {\bibinfo  {journal} {Nat. Mater.}\
  }\textbf {\bibinfo {volume} {11}},\ \bibinfo {pages} {1013} (\bibinfo {year}
  {2012})}\BibitemShut {NoStop}%
\bibitem [{\citenamefont {Watanabe}\ and\ \citenamefont
  {Miyake}(2013)}]{watanabe13}%
  \BibitemOpen
  \bibfield  {author} {\bibinfo {author} {\bibfnamefont {S.}~\bibnamefont
  {Watanabe}}\ and\ \bibinfo {author} {\bibfnamefont {K.}~\bibnamefont
  {Miyake}},\ }\bibfield  {title} {\enquote {\bibinfo {title} {{Robustness of
  Quantum Criticality of Valence Fluctuations}},}\ }\href {\doibase
  10.7566/JPSJ.82.083704} {\bibfield  {journal} {\bibinfo  {journal} {J. Phys.
  Soc. Jpn.}\ }\textbf {\bibinfo {volume} {82}},\ \bibinfo {pages} {083704}
  (\bibinfo {year} {2013})}\BibitemShut {NoStop}%
\bibitem [{\citenamefont {Shaginyan}\ \emph {et~al.}(2013)\citenamefont
  {Shaginyan}, \citenamefont {Msezane}, \citenamefont {Popov}, \citenamefont
  {Japaridze},\ and\ \citenamefont {Khodel}}]{shaginyan13}%
  \BibitemOpen
  \bibfield  {author} {\bibinfo {author} {\bibfnamefont {V.~R.}\ \bibnamefont
  {Shaginyan}}, \bibinfo {author} {\bibfnamefont {A.~Z.}\ \bibnamefont
  {Msezane}}, \bibinfo {author} {\bibfnamefont {K.~G.}\ \bibnamefont {Popov}},
  \bibinfo {author} {\bibfnamefont {G.~S.}\ \bibnamefont {Japaridze}}, \ and\
  \bibinfo {author} {\bibfnamefont {V.~A.}\ \bibnamefont {Khodel}},\ }\bibfield
   {title} {\enquote {\bibinfo {title} {{Common quantum phase transition in
  quasicrystals and heavy-fermion metals}},}\ }\href {\doibase
  10.1103/PhysRevB.87.245122} {\bibfield  {journal} {\bibinfo  {journal} {Phys.
  Rev. B}\ }\textbf {\bibinfo {volume} {87}},\ \bibinfo {pages} {245122}
  (\bibinfo {year} {2013})}\BibitemShut {NoStop}%
\bibitem [{\citenamefont {Andrade}\ \emph {et~al.}(2015)\citenamefont
  {Andrade}, \citenamefont {Jagannathan}, \citenamefont {Miranda},
  \citenamefont {Vojta},\ and\ \citenamefont {Dobrosavljevi\'{c}}}]{andrade15}%
  \BibitemOpen
  \bibfield  {author} {\bibinfo {author} {\bibfnamefont {E.~C.}\ \bibnamefont
  {Andrade}}, \bibinfo {author} {\bibfnamefont {A.}~\bibnamefont
  {Jagannathan}}, \bibinfo {author} {\bibfnamefont {E.}~\bibnamefont
  {Miranda}}, \bibinfo {author} {\bibfnamefont {M.}~\bibnamefont {Vojta}}, \
  and\ \bibinfo {author} {\bibfnamefont {V.}~\bibnamefont
  {Dobrosavljevi\'{c}}},\ }\bibfield  {title} {\enquote {\bibinfo {title}
  {{Non-Fermi-Liquid Behavior in Metallic Quasicrystals with Local Magnetic
  Moments}},}\ }\href {\doibase 10.1103/PhysRevLett.115.036403} {\bibfield
  {journal} {\bibinfo  {journal} {Phys. Rev. Lett.}\ }\textbf {\bibinfo
  {volume} {115}},\ \bibinfo {pages} {036403} (\bibinfo {year}
  {2015})}\BibitemShut {NoStop}%
\bibitem [{\citenamefont {Takemura}\ \emph {et~al.}(2015)\citenamefont
  {Takemura}, \citenamefont {Takemori},\ and\ \citenamefont
  {Koga}}]{takemura15}%
  \BibitemOpen
  \bibfield  {author} {\bibinfo {author} {\bibfnamefont {S.}~\bibnamefont
  {Takemura}}, \bibinfo {author} {\bibfnamefont {N.}~\bibnamefont {Takemori}},
  \ and\ \bibinfo {author} {\bibfnamefont {A.}~\bibnamefont {Koga}},\
  }\bibfield  {title} {\enquote {\bibinfo {title} {{Valence fluctuations and
  electric reconstruction in the extended Anderson model on the two-dimensional
  Penrose lattice}},}\ }\href {\doibase 10.1103/PhysRevB.91.165114} {\bibfield
  {journal} {\bibinfo  {journal} {Phys. Rev. B}\ }\textbf {\bibinfo {volume}
  {91}},\ \bibinfo {pages} {165114} (\bibinfo {year} {2015})}\BibitemShut
  {NoStop}%
\bibitem [{\citenamefont {Graebner}\ and\ \citenamefont
  {Chen}(1987)}]{graebner87}%
  \BibitemOpen
  \bibfield  {author} {\bibinfo {author} {\bibfnamefont {J.~E.}\ \bibnamefont
  {Graebner}}\ and\ \bibinfo {author} {\bibfnamefont {H.~S.}\ \bibnamefont
  {Chen}},\ }\bibfield  {title} {\enquote {\bibinfo {title} {{Specific Heat of
  an Icosahedral Superconductor, Mg$_{3}$Zn$_{3}$Al$_{2}$}},}\ }\href {\doibase
  10.1103/PhysRevLett.58.1945} {\bibfield  {journal} {\bibinfo  {journal}
  {Phys. Rev. Lett.}\ }\textbf {\bibinfo {volume} {58}},\ \bibinfo {pages}
  {1945} (\bibinfo {year} {1987})}\BibitemShut {NoStop}%
\bibitem [{\citenamefont {Deguchi}\ \emph {et~al.}(2015)\citenamefont
  {Deguchi}, \citenamefont {Nakayama}, \citenamefont {Matsukawa}, \citenamefont
  {Imura}, \citenamefont {Tanaka}, \citenamefont {Ishimasa},\ and\
  \citenamefont {Sato}}]{deguchi15}%
  \BibitemOpen
  \bibfield  {author} {\bibinfo {author} {\bibfnamefont {K.}~\bibnamefont
  {Deguchi}}, \bibinfo {author} {\bibfnamefont {M.}~\bibnamefont {Nakayama}},
  \bibinfo {author} {\bibfnamefont {S.}~\bibnamefont {Matsukawa}}, \bibinfo
  {author} {\bibfnamefont {K.}~\bibnamefont {Imura}}, \bibinfo {author}
  {\bibfnamefont {K.}~\bibnamefont {Tanaka}}, \bibinfo {author} {\bibfnamefont
  {T.}~\bibnamefont {Ishimasa}}, \ and\ \bibinfo {author} {\bibfnamefont
  {N.~K.}\ \bibnamefont {Sato}},\ }\bibfield  {title} {\enquote {\bibinfo
  {title} {{Superconductivity of Au-Ge-Yb Approximants with Tsai-Type
  Clusters}},}\ }\href {\doibase 10.7566/JPSJ.84.023705} {\bibfield  {journal}
  {\bibinfo  {journal} {J. Phys. Soc. Jpn.}\ }\textbf {\bibinfo {volume}
  {84}},\ \bibinfo {pages} {023705} (\bibinfo {year} {2015})}\BibitemShut
  {NoStop}%
\bibitem [{\citenamefont {Kamiya}\ \emph {et~al.}(2018)\citenamefont {Kamiya},
  \citenamefont {Takeuchi}, \citenamefont {Kabeya}, \citenamefont {Wada},
  \citenamefont {Ishimasa}, \citenamefont {Ochiai}, \citenamefont {Deguchi},
  \citenamefont {Imura},\ and\ \citenamefont {Sato}}]{kamiya18}%
  \BibitemOpen
  \bibfield  {author} {\bibinfo {author} {\bibfnamefont {K.}~\bibnamefont
  {Kamiya}}, \bibinfo {author} {\bibfnamefont {T.}~\bibnamefont {Takeuchi}},
  \bibinfo {author} {\bibfnamefont {N.}~\bibnamefont {Kabeya}}, \bibinfo
  {author} {\bibfnamefont {N.}~\bibnamefont {Wada}}, \bibinfo {author}
  {\bibfnamefont {T.}~\bibnamefont {Ishimasa}}, \bibinfo {author}
  {\bibfnamefont {A.}~\bibnamefont {Ochiai}}, \bibinfo {author} {\bibfnamefont
  {K.}~\bibnamefont {Deguchi}}, \bibinfo {author} {\bibfnamefont
  {K.}~\bibnamefont {Imura}}, \ and\ \bibinfo {author} {\bibfnamefont {N.~K.}\
  \bibnamefont {Sato}},\ }\bibfield  {title} {\enquote {\bibinfo {title}
  {{Discovery of superconductivity in quasicrystal}},}\ }\href {\doibase
  10.1038/s41467-017-02667-x} {\bibfield  {journal} {\bibinfo  {journal} {Nat.
  Commun.}\ }\textbf {\bibinfo {volume} {9}},\ \bibinfo {pages} {154} (\bibinfo
  {year} {2018})}\BibitemShut {NoStop}%
\bibitem [{lcd()}]{lcd}%
  \BibitemOpen
  \href@noop {} {}\bibinfo {note} {As argued in Ref.~\onlinecite{luck86b}, the
  lower critical dimension for the existence of critical states in
  quasicrystals is $1D$. In this sense, there should not be significant
  qualitative differences between electronic quasicrystalline states in $2D$
  and $3D$.}\BibitemShut {Stop}%
\bibitem [{\citenamefont {Socolar}(1989)}]{socolar89}%
  \BibitemOpen
  \bibfield  {author} {\bibinfo {author} {\bibfnamefont {J.~E.~S.}\
  \bibnamefont {Socolar}},\ }\bibfield  {title} {\enquote {\bibinfo {title}
  {{Simple octagonal and dodecagonal quasicrystals}},}\ }\href {\doibase
  10.1103/PhysRevB.39.10519} {\bibfield  {journal} {\bibinfo  {journal} {Phys.
  Rev. B}\ }\textbf {\bibinfo {volume} {39}},\ \bibinfo {pages} {10519}
  (\bibinfo {year} {1989})}\BibitemShut {NoStop}%
\bibitem [{\citenamefont {Levine}\ and\ \citenamefont
  {Steinhardt}(1987)}]{levine87}%
  \BibitemOpen
  \bibfield  {author} {\bibinfo {author} {\bibfnamefont {D.}~\bibnamefont
  {Levine}}\ and\ \bibinfo {author} {\bibfnamefont {P.~J.}\ \bibnamefont
  {Steinhardt}},\ }\href@noop {} {\emph {\bibinfo {title} {{The Physics of
  Quasicrystals}}}}\ (\bibinfo  {publisher} {World Scientific, Singapore},\
  \bibinfo {year} {1987})\BibitemShut {NoStop}%
\bibitem [{\citenamefont {Duneau}\ \emph {et~al.}(1989)\citenamefont {Duneau},
  \citenamefont {Mosseri},\ and\ \citenamefont {Oguey}}]{deneau89}%
  \BibitemOpen
  \bibfield  {author} {\bibinfo {author} {\bibfnamefont {M.}~\bibnamefont
  {Duneau}}, \bibinfo {author} {\bibfnamefont {R.}~\bibnamefont {Mosseri}}, \
  and\ \bibinfo {author} {\bibfnamefont {C.}~\bibnamefont {Oguey}},\ }\bibfield
   {title} {\enquote {\bibinfo {title} {{Approximants of quasiperiodic
  structures generated by the inflation mapping}},}\ }\href@noop {} {\bibfield
  {journal} {\bibinfo  {journal} {J. Phys. A: Math. Gen.}\ }\textbf {\bibinfo
  {volume} {22}},\ \bibinfo {pages} {4549} (\bibinfo {year}
  {1989})}\BibitemShut {NoStop}%
\bibitem [{\citenamefont {Benza}\ and\ \citenamefont {Sire}(1991)}]{benza91}%
  \BibitemOpen
  \bibfield  {author} {\bibinfo {author} {\bibfnamefont {V.~G.}\ \bibnamefont
  {Benza}}\ and\ \bibinfo {author} {\bibfnamefont {C.}~\bibnamefont {Sire}},\
  }\bibfield  {title} {\enquote {\bibinfo {title} {{Band spectrum of the
  octagonal quasicrystal: Finite measure, gaps, and chaos}},}\ }\href {\doibase
  10.1103/PhysRevB.44.10343} {\bibfield  {journal} {\bibinfo  {journal} {Phys.
  Rev. B}\ }\textbf {\bibinfo {volume} {44}},\ \bibinfo {pages} {10343}
  (\bibinfo {year} {1991})}\BibitemShut {NoStop}%
\bibitem [{\citenamefont {Jagannathan}(2000)}]{anu00}%
  \BibitemOpen
  \bibfield  {author} {\bibinfo {author} {\bibfnamefont {A.}~\bibnamefont
  {Jagannathan}},\ }\bibfield  {title} {\enquote {\bibinfo {title}
  {{Self-similarity under inflation and level statistics: A study in two
  dimensions}},}\ }\href {\doibase 10.1103/PhysRevB.61.R834} {\bibfield
  {journal} {\bibinfo  {journal} {Phys. Rev. B}\ }\textbf {\bibinfo {volume}
  {61}},\ \bibinfo {pages} {R834} (\bibinfo {year} {2000})}\BibitemShut
  {NoStop}%
\bibitem [{\citenamefont {Smith}\ and\ \citenamefont
  {Ashcroft}(1987)}]{smith87}%
  \BibitemOpen
  \bibfield  {author} {\bibinfo {author} {\bibfnamefont {A.~P.}\ \bibnamefont
  {Smith}}\ and\ \bibinfo {author} {\bibfnamefont {N.~W.}\ \bibnamefont
  {Ashcroft}},\ }\bibfield  {title} {\enquote {\bibinfo {title}
  {{Pseudopotentials and quasicrystals}},}\ }\href {\doibase
  10.1103/PhysRevLett.59.1365} {\bibfield  {journal} {\bibinfo  {journal}
  {Phys. Rev. Lett.}\ }\textbf {\bibinfo {volume} {59}},\ \bibinfo {pages}
  {1365} (\bibinfo {year} {1987})}\BibitemShut {NoStop}%
\bibitem [{\citenamefont {Zijlstra}\ and\ \citenamefont
  {Janssen}(2000)}]{zijlstra00}%
  \BibitemOpen
  \bibfield  {author} {\bibinfo {author} {\bibfnamefont {E.~S.}\ \bibnamefont
  {Zijlstra}}\ and\ \bibinfo {author} {\bibfnamefont {T.}~\bibnamefont
  {Janssen}},\ }\bibfield  {title} {\enquote {\bibinfo {title} {{Non-spiky
  density of states of an icosahedral quasicrystal}},}\ }\href {\doibase
  10.1209/epl/i2000-00476-x} {\bibfield  {journal} {\bibinfo  {journal}
  {Europhys. Lett.}\ }\textbf {\bibinfo {volume} {52}},\ \bibinfo {pages} {578}
  (\bibinfo {year} {2000})}\BibitemShut {NoStop}%
\bibitem [{\citenamefont {Richardella}\ \emph {et~al.}(2010)\citenamefont
  {Richardella}, \citenamefont {Roushan}, \citenamefont {Mack}, \citenamefont
  {Zhou}, \citenamefont {Huse}, \citenamefont {Awschalom},\ and\ \citenamefont
  {Yazdani}}]{richardella10}%
  \BibitemOpen
  \bibfield  {author} {\bibinfo {author} {\bibfnamefont {A.}~\bibnamefont
  {Richardella}}, \bibinfo {author} {\bibfnamefont {P.}~\bibnamefont
  {Roushan}}, \bibinfo {author} {\bibfnamefont {S.}~\bibnamefont {Mack}},
  \bibinfo {author} {\bibfnamefont {B.}~\bibnamefont {Zhou}}, \bibinfo {author}
  {\bibfnamefont {D.~A.}\ \bibnamefont {Huse}}, \bibinfo {author}
  {\bibfnamefont {D.~D.}\ \bibnamefont {Awschalom}}, \ and\ \bibinfo {author}
  {\bibfnamefont {A.}~\bibnamefont {Yazdani}},\ }\bibfield  {title} {\enquote
  {\bibinfo {title} {{Visualizing Critical Correlations Near the
  Metal-Insulator Transition in Ga$_{1-x}$Mn$_{x}$As}},}\ }\href {\doibase
  10.1126/science.1183640} {\bibfield  {journal} {\bibinfo  {journal}
  {Science}\ }\textbf {\bibinfo {volume} {327}},\ \bibinfo {pages} {665}
  (\bibinfo {year} {2010})}\BibitemShut {NoStop}%
\bibitem [{\citenamefont {Rodriguez}\ \emph {et~al.}(2010)\citenamefont
  {Rodriguez}, \citenamefont {Vasquez}, \citenamefont {Slevin},\ and\
  \citenamefont {R\"{o}mer}}]{rodriguez10}%
  \BibitemOpen
  \bibfield  {author} {\bibinfo {author} {\bibfnamefont {A.}~\bibnamefont
  {Rodriguez}}, \bibinfo {author} {\bibfnamefont {L.~J.}\ \bibnamefont
  {Vasquez}}, \bibinfo {author} {\bibfnamefont {K.}~\bibnamefont {Slevin}}, \
  and\ \bibinfo {author} {\bibfnamefont {R.~A.}\ \bibnamefont {R\"{o}mer}},\
  }\bibfield  {title} {\enquote {\bibinfo {title} {{Critical Parameters from a
  Generalized Multifractal Analysis at the Anderson Transition}},}\ }\href
  {\doibase 10.1103/PhysRevLett.105.046403} {\bibfield  {journal} {\bibinfo
  {journal} {Phys. Rev. Lett.}\ }\textbf {\bibinfo {volume} {105}},\ \bibinfo
  {pages} {046403} (\bibinfo {year} {2010})}\BibitemShut {NoStop}%
\bibitem [{\citenamefont {Rodriguez}\ \emph {et~al.}(2011)\citenamefont
  {Rodriguez}, \citenamefont {Vasquez}, \citenamefont {Slevin},\ and\
  \citenamefont {R\"{o}mer}}]{rodriguez11}%
  \BibitemOpen
  \bibfield  {author} {\bibinfo {author} {\bibfnamefont {A.}~\bibnamefont
  {Rodriguez}}, \bibinfo {author} {\bibfnamefont {L.~J.}\ \bibnamefont
  {Vasquez}}, \bibinfo {author} {\bibfnamefont {K.}~\bibnamefont {Slevin}}, \
  and\ \bibinfo {author} {\bibfnamefont {R.~A.}\ \bibnamefont {R\"{o}mer}},\
  }\bibfield  {title} {\enquote {\bibinfo {title} {{Multifractal finite-size
  scaling and universality at the Anderson transition}},}\ }\href {\doibase
  10.1103/PhysRevB.84.134209} {\bibfield  {journal} {\bibinfo  {journal} {Phys.
  Rev. B}\ }\textbf {\bibinfo {volume} {84}},\ \bibinfo {pages} {134209}
  (\bibinfo {year} {2011})}\BibitemShut {NoStop}%
\bibitem [{\citenamefont {Stadnik}\ \emph {et~al.}(1997)\citenamefont
  {Stadnik}, \citenamefont {Purdie}, \citenamefont {Garnier}, \citenamefont
  {Baer}, \citenamefont {Tsai}, \citenamefont {Inoue}, \citenamefont {Edagawa},
  \citenamefont {Takeuchi},\ and\ \citenamefont {Buschow}}]{stadinik97}%
  \BibitemOpen
  \bibfield  {author} {\bibinfo {author} {\bibfnamefont {Z.~M.}\ \bibnamefont
  {Stadnik}}, \bibinfo {author} {\bibfnamefont {D.}~\bibnamefont {Purdie}},
  \bibinfo {author} {\bibfnamefont {M.}~\bibnamefont {Garnier}}, \bibinfo
  {author} {\bibfnamefont {Y.}~\bibnamefont {Baer}}, \bibinfo {author}
  {\bibfnamefont {A.-P.}\ \bibnamefont {Tsai}}, \bibinfo {author}
  {\bibfnamefont {A.}~\bibnamefont {Inoue}}, \bibinfo {author} {\bibfnamefont
  {K.}~\bibnamefont {Edagawa}}, \bibinfo {author} {\bibfnamefont
  {S.}~\bibnamefont {Takeuchi}}, \ and\ \bibinfo {author} {\bibfnamefont
  {K.~H.~J.}\ \bibnamefont {Buschow}},\ }\bibfield  {title} {\enquote {\bibinfo
  {title} {{Electronic structure of quasicrystals studied by
  ultrahigh-energy-resolution photoemission spectroscopy}},}\ }\href {\doibase
  10.1103/PhysRevB.55.10938} {\bibfield  {journal} {\bibinfo  {journal} {Phys.
  Rev. B}\ }\textbf {\bibinfo {volume} {55}},\ \bibinfo {pages} {10938}
  (\bibinfo {year} {1997})}\BibitemShut {NoStop}%
\bibitem [{\citenamefont {Rotenberg}\ \emph {et~al.}()\citenamefont
  {Rotenberg}, \citenamefont {Theis}, \citenamefont {Horn},\ and\ \citenamefont
  {Gille}}]{rotengerg00}%
  \BibitemOpen
  \bibfield  {author} {\bibinfo {author} {\bibfnamefont {E.}~\bibnamefont
  {Rotenberg}}, \bibinfo {author} {\bibfnamefont {W.}~\bibnamefont {Theis}},
  \bibinfo {author} {\bibfnamefont {K.}~\bibnamefont {Horn}}, \ and\ \bibinfo
  {author} {\bibfnamefont {P.}~\bibnamefont {Gille}},\ }\bibfield  {title}
  {\enquote {\bibinfo {title} {{Quasicrystalline valence bands in decagonal
  AlNiCo}},}\ }\href {\doibase 10.1038/35020519} {\bibfield  {journal}
  {\bibinfo  {journal} {Nature}\ }\textbf {\bibinfo {volume} {406}},\ \bibinfo
  {pages} {602}}\BibitemShut {NoStop}%
\bibitem [{\citenamefont {Rogalev}\ \emph {et~al.}(2015)\citenamefont
  {Rogalev}, \citenamefont {Gr\"{o}ning}, \citenamefont {Widmer}, \citenamefont
  {Dil}, \citenamefont {Bisti}, \citenamefont {Lev}, \citenamefont {Schmitt},\
  and\ \citenamefont {Strocov}}]{rogalev15}%
  \BibitemOpen
  \bibfield  {author} {\bibinfo {author} {\bibfnamefont {V.~A.}\ \bibnamefont
  {Rogalev}}, \bibinfo {author} {\bibfnamefont {O.}~\bibnamefont
  {Gr\"{o}ning}}, \bibinfo {author} {\bibfnamefont {R.}~\bibnamefont {Widmer}},
  \bibinfo {author} {\bibfnamefont {J.~H.}\ \bibnamefont {Dil}}, \bibinfo
  {author} {\bibfnamefont {F.}~\bibnamefont {Bisti}}, \bibinfo {author}
  {\bibfnamefont {L.~L.}\ \bibnamefont {Lev}}, \bibinfo {author} {\bibfnamefont
  {T.}~\bibnamefont {Schmitt}}, \ and\ \bibinfo {author} {\bibfnamefont
  {V.~N.}\ \bibnamefont {Strocov}},\ }\bibfield  {title} {\enquote {\bibinfo
  {title} {{Fermi states and anisotropy of Brillouin zone scattering in the
  decagonal Al-Ni-Co quasicrystal}},}\ }\href {\doibase
  10.1038/ncomms9607;;;;;; 10.1038/ncomms9607} {\bibfield  {journal} {\bibinfo
  {journal} {Nat. Commun.}\ }\textbf {\bibinfo {volume} {6}},\ \bibinfo {pages}
  {8607} (\bibinfo {year} {2015})}\BibitemShut {NoStop}%
\bibitem [{\citenamefont {Chhabra}\ and\ \citenamefont
  {Jensen}(1989)}]{chhabra89}%
  \BibitemOpen
  \bibfield  {author} {\bibinfo {author} {\bibfnamefont {A.}~\bibnamefont
  {Chhabra}}\ and\ \bibinfo {author} {\bibfnamefont {R.~V.}\ \bibnamefont
  {Jensen}},\ }\bibfield  {title} {\enquote {\bibinfo {title} {{Direct
  determination of the f($\alpha${}) singularity spectrum}},}\ }\href {\doibase
  10.1103/PhysRevLett.62.1327} {\bibfield  {journal} {\bibinfo  {journal}
  {Phys. Rev. Lett.}\ }\textbf {\bibinfo {volume} {62}},\ \bibinfo {pages}
  {1327} (\bibinfo {year} {1989})}\BibitemShut {NoStop}%
\bibitem [{\citenamefont {Varma}\ \emph {et~al.}(2016)\citenamefont {Varma},
  \citenamefont {Pilati},\ and\ \citenamefont {Kravtsov}}]{varma16}%
  \BibitemOpen
  \bibfield  {author} {\bibinfo {author} {\bibfnamefont {V.~K.}\ \bibnamefont
  {Varma}}, \bibinfo {author} {\bibfnamefont {S.}~\bibnamefont {Pilati}}, \
  and\ \bibinfo {author} {\bibfnamefont {V.~E.}\ \bibnamefont {Kravtsov}},\
  }\bibfield  {title} {\enquote {\bibinfo {title} {{Conduction in quasiperiodic
  and quasirandom lattices: Fibonacci, Riemann, and Anderson models}},}\ }\href
  {\doibase 10.1103/PhysRevB.94.214204} {\bibfield  {journal} {\bibinfo
  {journal} {Phys. Rev. B}\ }\textbf {\bibinfo {volume} {94}},\ \bibinfo
  {pages} {214204} (\bibinfo {year} {2016})}\BibitemShut {NoStop}%
\bibitem [{\citenamefont {Resta}\ and\ \citenamefont
  {Sorella}(1999)}]{resta99}%
  \BibitemOpen
  \bibfield  {author} {\bibinfo {author} {\bibfnamefont {R.}~\bibnamefont
  {Resta}}\ and\ \bibinfo {author} {\bibfnamefont {S.}~\bibnamefont
  {Sorella}},\ }\bibfield  {title} {\enquote {\bibinfo {title} {{Electron
  Localization in the Insulating State}},}\ }\href {\doibase
  10.1103/PhysRevLett.82.370} {\bibfield  {journal} {\bibinfo  {journal} {Phys.
  Rev. Lett.}\ }\textbf {\bibinfo {volume} {82}},\ \bibinfo {pages} {370}
  (\bibinfo {year} {1999})}\BibitemShut {NoStop}%
\bibitem [{\citenamefont {Resta}(2011)}]{resta11}%
  \BibitemOpen
  \bibfield  {author} {\bibinfo {author} {\bibfnamefont {R.}~\bibnamefont
  {Resta}},\ }\bibfield  {title} {\enquote {\bibinfo {title} {{The insulating
  state of matter: a geometrical theory}},}\ }\href {\doibase
  10.1140/epjb/e2010-10874-4} {\bibfield  {journal} {\bibinfo  {journal} {Eur.
  Phys. J. B}\ }\textbf {\bibinfo {volume} {79}},\ \bibinfo {pages} {121}
  (\bibinfo {year} {2011})}\BibitemShut {NoStop}%
\bibitem [{\citenamefont {Souza}\ \emph {et~al.}()\citenamefont {Souza},
  \citenamefont {Wilkens},\ and\ \citenamefont {Martin}}]{souza00}%
  \BibitemOpen
  \bibfield  {author} {\bibinfo {author} {\bibfnamefont {I.}~\bibnamefont
  {Souza}}, \bibinfo {author} {\bibfnamefont {T.}~\bibnamefont {Wilkens}}, \
  and\ \bibinfo {author} {\bibfnamefont {R.~M.}\ \bibnamefont {Martin}},\
  }\bibfield  {title} {\enquote {\bibinfo {title} {{Polarization and
  localization in insulators: Generating function approach}},}\ }\href
  {\doibase 10.1103/PhysRevB.62.1666} {\bibfield  {journal} {\bibinfo
  {journal} {Phys. Rev. B}\ }\textbf {\bibinfo {volume} {62}},\ \bibinfo
  {pages} {1666} (\bibinfo {year} {2000})}\BibitemShut {NoStop}%
\bibitem [{\citenamefont {Luck}\ and\ \citenamefont
  {Petritis}(1986)}]{luck86b}%
  \BibitemOpen
  \bibfield  {author} {\bibinfo {author} {\bibfnamefont {J.~M.}\ \bibnamefont
  {Luck}}\ and\ \bibinfo {author} {\bibfnamefont {D.}~\bibnamefont
  {Petritis}},\ }\bibfield  {title} {\enquote {\bibinfo {title} {{Phonon
  spectra in one-dimensional quasicrystals}},}\ }\href {\doibase
  10.1007/BF01127714} {\bibfield  {journal} {\bibinfo  {journal} {J. Stat.
  Phys.}\ }\textbf {\bibinfo {volume} {42}},\ \bibinfo {pages} {289} (\bibinfo
  {year} {1986})}\BibitemShut {NoStop}%
\bibitem [{\citenamefont {Los}\ \emph {et~al.}(1993)\citenamefont {Los},
  \citenamefont {Janssen},\ and\ \citenamefont {G\"{a}hler}}]{los93}%
  \BibitemOpen
  \bibfield  {author} {\bibinfo {author} {\bibfnamefont {J.}~\bibnamefont
  {Los}}, \bibinfo {author} {\bibfnamefont {T.}~\bibnamefont {Janssen}}, \ and\
  \bibinfo {author} {\bibfnamefont {F.}~\bibnamefont {G\"{a}hler}},\ }\bibfield
   {title} {\enquote {\bibinfo {title} {{The Phonon Spectrum of the octagonal
  tiling}},}\ }\href {\doibase 10.1142/S0217979293002468} {\bibfield  {journal}
  {\bibinfo  {journal} {Int. J. Mod. Phys. B}\ }\textbf {\bibinfo {volume}
  {07}},\ \bibinfo {pages} {1505} (\bibinfo {year} {1993})}\BibitemShut
  {NoStop}%
\bibitem [{\citenamefont {Quilichini}(1997)}]{quilichini97}%
  \BibitemOpen
  \bibfield  {author} {\bibinfo {author} {\bibfnamefont {M.}~\bibnamefont
  {Quilichini}},\ }\bibfield  {title} {\enquote {\bibinfo {title} {{Phonon
  excitations in quasicrystals}},}\ }\href {\doibase 10.1103/RevModPhys.69.277}
  {\bibfield  {journal} {\bibinfo  {journal} {Rev. Mod. Phys.}\ }\textbf
  {\bibinfo {volume} {69}},\ \bibinfo {pages} {277} (\bibinfo {year}
  {1997})}\BibitemShut {NoStop}%
\bibitem [{\citenamefont {de~Boissieu}(2012)}]{deboissieu12}%
  \BibitemOpen
  \bibfield  {author} {\bibinfo {author} {\bibfnamefont {M.}~\bibnamefont
  {de~Boissieu}},\ }\bibfield  {title} {\enquote {\bibinfo {title} {{Phonons{,}
  phasons and atomic dynamics in quasicrystals}},}\ }\href {\doibase
  10.1039/C2CS35212E} {\bibfield  {journal} {\bibinfo  {journal} {Chem. Soc.
  Rev.}\ }\textbf {\bibinfo {volume} {41}},\ \bibinfo {pages} {6778} (\bibinfo
  {year} {2012})}\BibitemShut {NoStop}%
\bibitem [{\citenamefont {Brown}\ \emph {et~al.}(2018)\citenamefont {Brown},
  \citenamefont {Semeniuk}, \citenamefont {Wang}, \citenamefont {Monserrat},
  \citenamefont {Pickard},\ and\ \citenamefont {Grosche}}]{brown18}%
  \BibitemOpen
  \bibfield  {author} {\bibinfo {author} {\bibfnamefont {P.}~\bibnamefont
  {Brown}}, \bibinfo {author} {\bibfnamefont {K.}~\bibnamefont {Semeniuk}},
  \bibinfo {author} {\bibfnamefont {D.}~\bibnamefont {Wang}}, \bibinfo {author}
  {\bibfnamefont {B.}~\bibnamefont {Monserrat}}, \bibinfo {author}
  {\bibfnamefont {C.~J.}\ \bibnamefont {Pickard}}, \ and\ \bibinfo {author}
  {\bibfnamefont {F.~M.}\ \bibnamefont {Grosche}},\ }\bibfield  {title}
  {\enquote {\bibinfo {title} {{Strong coupling superconductivity in a
  quasiperiodic host-guest structure}},}\ }\href {\doibase
  10.1126/sciadv.aao4793} {\bibfield  {journal} {\bibinfo  {journal} {Sci.
  Adv.}\ }\textbf {\bibinfo {volume} {4}} (\bibinfo {year} {2018}),\
  10.1126/sciadv.aao4793}\BibitemShut {NoStop}%
\bibitem [{\citenamefont {Sakai}\ \emph {et~al.}(2017)\citenamefont {Sakai},
  \citenamefont {Takemori}, \citenamefont {Koga},\ and\ \citenamefont
  {Arita}}]{sakai17}%
  \BibitemOpen
  \bibfield  {author} {\bibinfo {author} {\bibfnamefont {S.}~\bibnamefont
  {Sakai}}, \bibinfo {author} {\bibfnamefont {N.}~\bibnamefont {Takemori}},
  \bibinfo {author} {\bibfnamefont {A.}~\bibnamefont {Koga}}, \ and\ \bibinfo
  {author} {\bibfnamefont {R.}~\bibnamefont {Arita}},\ }\bibfield  {title}
  {\enquote {\bibinfo {title} {{Superconductivity on a quasiperiodic lattice:
  Extended-to-localized crossover of Cooper pairs}},}\ }\href {\doibase
  10.1103/PhysRevB.95.024509} {\bibfield  {journal} {\bibinfo  {journal} {Phys.
  Rev. B}\ }\textbf {\bibinfo {volume} {95}},\ \bibinfo {pages} {024509}
  (\bibinfo {year} {2017})}\BibitemShut {NoStop}%
\bibitem [{\citenamefont {Ghosal}\ \emph {et~al.}(1998)\citenamefont {Ghosal},
  \citenamefont {Randeria},\ and\ \citenamefont {Trivedi}}]{ghosal98}%
  \BibitemOpen
  \bibfield  {author} {\bibinfo {author} {\bibfnamefont {A.}~\bibnamefont
  {Ghosal}}, \bibinfo {author} {\bibfnamefont {M.}~\bibnamefont {Randeria}}, \
  and\ \bibinfo {author} {\bibfnamefont {N.}~\bibnamefont {Trivedi}},\
  }\bibfield  {title} {\enquote {\bibinfo {title} {{Role of Spatial Amplitude
  Fluctuations in Highly Disordered $\mathit{s}$-Wave Superconductors}},}\
  }\href {\doibase 10.1103/PhysRevLett.81.3940} {\bibfield  {journal} {\bibinfo
   {journal} {Phys. Rev. Lett.}\ }\textbf {\bibinfo {volume} {81}},\ \bibinfo
  {pages} {3940} (\bibinfo {year} {1998})}\BibitemShut {NoStop}%
\bibitem [{\citenamefont {Ghosal}\ \emph {et~al.}(2001)\citenamefont {Ghosal},
  \citenamefont {Randeria},\ and\ \citenamefont {Trivedi}}]{ghosal01}%
  \BibitemOpen
  \bibfield  {author} {\bibinfo {author} {\bibfnamefont {A.}~\bibnamefont
  {Ghosal}}, \bibinfo {author} {\bibfnamefont {M.}~\bibnamefont {Randeria}}, \
  and\ \bibinfo {author} {\bibfnamefont {N.}~\bibnamefont {Trivedi}},\
  }\bibfield  {title} {\enquote {\bibinfo {title} {{Inhomogeneous pairing in
  highly disordered s-wave superconductors}},}\ }\href {\doibase
  10.1103/PhysRevB.65.014501} {\bibfield  {journal} {\bibinfo  {journal} {Phys.
  Rev. B}\ }\textbf {\bibinfo {volume} {65}},\ \bibinfo {pages} {014501}
  (\bibinfo {year} {2001})}\BibitemShut {NoStop}%
\bibitem [{\citenamefont {Dobrosavljevi\'{c}}\ \emph
  {et~al.}(2012)\citenamefont {Dobrosavljevi\'{c}}, \citenamefont {Trivedi},\
  and\ \citenamefont {{Valles Jr}}}]{vlad_book}%
  \BibitemOpen
  \bibfield  {author} {\bibinfo {author} {\bibfnamefont {V.}~\bibnamefont
  {Dobrosavljevi\'{c}}}, \bibinfo {author} {\bibfnamefont {N.}~\bibnamefont
  {Trivedi}}, \ and\ \bibinfo {author} {\bibfnamefont {J.~M.}\ \bibnamefont
  {{Valles Jr}}},\ }\href@noop {} {\emph {\bibinfo {title} {{Conductor
  Insulator Quantum Phase Transitions}}}}\ (\bibinfo  {publisher} {Oxford
  University Press},\ \bibinfo {address} {UK},\ \bibinfo {year}
  {2012})\BibitemShut {NoStop}%
\bibitem [{\citenamefont {Collins}\ \emph {et~al.}(2017)\citenamefont
  {Collins}, \citenamefont {Witte}, \citenamefont {Silverman}, \citenamefont
  {Green},\ and\ \citenamefont {K.}}]{collins17}%
  \BibitemOpen
  \bibfield  {author} {\bibinfo {author} {\bibfnamefont {L.~C.}\ \bibnamefont
  {Collins}}, \bibinfo {author} {\bibfnamefont {T.~G.}\ \bibnamefont {Witte}},
  \bibinfo {author} {\bibfnamefont {R.}~\bibnamefont {Silverman}}, \bibinfo
  {author} {\bibfnamefont {D.~B.}\ \bibnamefont {Green}}, \ and\ \bibinfo
  {author} {\bibfnamefont {Gomes~K.}\ \bibnamefont {K.}},\ }\bibfield  {title}
  {\enquote {\bibinfo {title} {{Imaging quasiperiodic electronic states in a
  synthetic Penrose tiling}},}\ }\href {\doibase 10.1038/ncomms15961;;;
  10.1038/ncomms15961} {\bibfield  {journal} {\bibinfo  {journal} {Nat.
  Commun.}\ }\textbf {\bibinfo {volume} {8}},\ \bibinfo {pages} {15961}
  (\bibinfo {year} {2017})}\BibitemShut {NoStop}%
\bibitem [{\citenamefont {Bouadim}\ \emph {et~al.}(2011)\citenamefont
  {Bouadim}, \citenamefont {Loh}, \citenamefont {Randeria},\ and\ \citenamefont
  {Trivedi}}]{bouadim11}%
  \BibitemOpen
  \bibfield  {author} {\bibinfo {author} {\bibfnamefont {K.}~\bibnamefont
  {Bouadim}}, \bibinfo {author} {\bibfnamefont {Y.~L.}\ \bibnamefont {Loh}},
  \bibinfo {author} {\bibfnamefont {M.}~\bibnamefont {Randeria}}, \ and\
  \bibinfo {author} {\bibfnamefont {N.}~\bibnamefont {Trivedi}},\ }\bibfield
  {title} {\enquote {\bibinfo {title} {{Single- and two-particle energy gaps
  across the disorder-driven superconductor-insulator transition}},}\ }\href
  {\doibase https://doi.org/10.1038/nphys2037} {\bibfield  {journal} {\bibinfo
  {journal} {Nature Physics}\ }\textbf {\bibinfo {volume} {7}},\ \bibinfo
  {pages} {884} (\bibinfo {year} {2011})}\BibitemShut {NoStop}%
\bibitem [{\citenamefont {Nandkishore}\ \emph {et~al.}(2013)\citenamefont
  {Nandkishore}, \citenamefont {Maciejko}, \citenamefont {Huse},\ and\
  \citenamefont {Sondhi}}]{nandkishore13}%
  \BibitemOpen
  \bibfield  {author} {\bibinfo {author} {\bibfnamefont {R.}~\bibnamefont
  {Nandkishore}}, \bibinfo {author} {\bibfnamefont {J.}~\bibnamefont
  {Maciejko}}, \bibinfo {author} {\bibfnamefont {D.~A.}\ \bibnamefont {Huse}},
  \ and\ \bibinfo {author} {\bibfnamefont {S.~L.}\ \bibnamefont {Sondhi}},\
  }\bibfield  {title} {\enquote {\bibinfo {title} {{Superconductivity of
  disordered Dirac fermions}},}\ }\href {\doibase 10.1103/PhysRevB.87.174511}
  {\bibfield  {journal} {\bibinfo  {journal} {Phys. Rev. B}\ }\textbf {\bibinfo
  {volume} {87}},\ \bibinfo {pages} {174511} (\bibinfo {year}
  {2013})}\BibitemShut {NoStop}%
\bibitem [{\citenamefont {Potirniche}\ \emph {et~al.}(2014)\citenamefont
  {Potirniche}, \citenamefont {Maciejko}, \citenamefont {Nandkishore},\ and\
  \citenamefont {Sondhi}}]{potirniche14}%
  \BibitemOpen
  \bibfield  {author} {\bibinfo {author} {\bibfnamefont {I.-D.}\ \bibnamefont
  {Potirniche}}, \bibinfo {author} {\bibfnamefont {J.}~\bibnamefont
  {Maciejko}}, \bibinfo {author} {\bibfnamefont {R.}~\bibnamefont
  {Nandkishore}}, \ and\ \bibinfo {author} {\bibfnamefont {S.~L.}\ \bibnamefont
  {Sondhi}},\ }\bibfield  {title} {\enquote {\bibinfo {title}
  {{Superconductivity of disordered Dirac fermions in graphene}},}\ }\href
  {\doibase 10.1103/PhysRevB.90.094516} {\bibfield  {journal} {\bibinfo
  {journal} {Phys. Rev. B}\ }\textbf {\bibinfo {volume} {90}},\ \bibinfo
  {pages} {094516} (\bibinfo {year} {2014})}\BibitemShut {NoStop}%
\bibitem [{\citenamefont {Dodaro}\ and\ \citenamefont
  {Kivelson}(2018)}]{dodaro18}%
  \BibitemOpen
  \bibfield  {author} {\bibinfo {author} {\bibfnamefont {J.~F.}\ \bibnamefont
  {Dodaro}}\ and\ \bibinfo {author} {\bibfnamefont {S.~A.}\ \bibnamefont
  {Kivelson}},\ }\bibfield  {title} {\enquote {\bibinfo {title}
  {{Generalization of Anderson's theorem for disordered superconductors}},}\
  }\href {\doibase 10.1103/PhysRevB.98.174503} {\bibfield  {journal} {\bibinfo
  {journal} {Phys. Rev. B}\ }\textbf {\bibinfo {volume} {98}},\ \bibinfo
  {pages} {174503} (\bibinfo {year} {2018})}\BibitemShut {NoStop}%
\bibitem [{\citenamefont {Andrade}\ \emph {et~al.}(2009)\citenamefont
  {Andrade}, \citenamefont {Miranda},\ and\ \citenamefont
  {Dobrosavljevi\'{c}}}]{amd09}%
  \BibitemOpen
  \bibfield  {author} {\bibinfo {author} {\bibfnamefont {E.~C.}\ \bibnamefont
  {Andrade}}, \bibinfo {author} {\bibfnamefont {E.}~\bibnamefont {Miranda}}, \
  and\ \bibinfo {author} {\bibfnamefont {V.}~\bibnamefont
  {Dobrosavljevi\'{c}}},\ }\bibfield  {title} {\enquote {\bibinfo {title}
  {{Electronic Griffiths Phase of the $d = 2$ Mott Transition}},}\ }\href
  {\doibase 10.1103/PhysRevLett.102.206403} {\bibfield  {journal} {\bibinfo
  {journal} {Phys. Rev. Lett.}\ }\textbf {\bibinfo {volume} {102}},\ \bibinfo
  {pages} {206403} (\bibinfo {year} {2009})}\BibitemShut {NoStop}%
\bibitem [{\citenamefont {Vojta}\ and\ \citenamefont {Hoyos}()}]{hoyos14}%
  \BibitemOpen
  \bibfield  {author} {\bibinfo {author} {\bibfnamefont {T.}~\bibnamefont
  {Vojta}}\ and\ \bibinfo {author} {\bibfnamefont {J.~A.}\ \bibnamefont
  {Hoyos}},\ }\bibfield  {title} {\enquote {\bibinfo {title} {{Criticality and
  Quenched Disorder: Harris Criterion Versus Rare Regions}},}\ }\href {\doibase
  10.1103/PhysRevLett.112.075702} {\bibfield  {journal} {\bibinfo  {journal}
  {Phys. Rev. Lett.}\ }\textbf {\bibinfo {volume} {112}},\ \bibinfo {pages}
  {075702}}\BibitemShut {NoStop}%
\bibitem [{\citenamefont {Feigel'man}\ \emph {et~al.}(2007)\citenamefont
  {Feigel'man}, \citenamefont {Ioffe}, \citenamefont {Kravtsov},\ and\
  \citenamefont {Yuzbashyan}}]{feigelman07}%
  \BibitemOpen
  \bibfield  {author} {\bibinfo {author} {\bibfnamefont {M.~V.}\ \bibnamefont
  {Feigel'man}}, \bibinfo {author} {\bibfnamefont {L.~B.}\ \bibnamefont
  {Ioffe}}, \bibinfo {author} {\bibfnamefont {V.~E.}\ \bibnamefont {Kravtsov}},
  \ and\ \bibinfo {author} {\bibfnamefont {E.~A.}\ \bibnamefont {Yuzbashyan}},\
  }\bibfield  {title} {\enquote {\bibinfo {title} {{Eigenfunction Fractality
  and Pseudogap State near the Superconductor-Insulator Transition}},}\ }\href
  {\doibase 10.1103/PhysRevLett.98.027001} {\bibfield  {journal} {\bibinfo
  {journal} {Phys. Rev. Lett.}\ }\textbf {\bibinfo {volume} {98}},\ \bibinfo
  {pages} {027001} (\bibinfo {year} {2007})}\BibitemShut {NoStop}%
\bibitem [{\citenamefont {Feigel'man}\ \emph {et~al.}(2010)\citenamefont
  {Feigel'man}, \citenamefont {Ioffe}, \citenamefont {Kravtsov},\ and\
  \citenamefont {Cuevas}}]{feigelman10}%
  \BibitemOpen
  \bibfield  {author} {\bibinfo {author} {\bibfnamefont {M.V.}\ \bibnamefont
  {Feigel'man}}, \bibinfo {author} {\bibfnamefont {L.B.}\ \bibnamefont
  {Ioffe}}, \bibinfo {author} {\bibfnamefont {V.E.}\ \bibnamefont {Kravtsov}},
  \ and\ \bibinfo {author} {\bibfnamefont {E.}~\bibnamefont {Cuevas}},\
  }\bibfield  {title} {\enquote {\bibinfo {title} {{Fractal superconductivity
  near localization threshold}},}\ }\href {\doibase 10.1016/j.aop.2010.04.001}
  {\bibfield  {journal} {\bibinfo  {journal} {Ann. Phys. (NY).}\ }\textbf
  {\bibinfo {volume} {325}},\ \bibinfo {pages} {1390} (\bibinfo {year}
  {2010})}\BibitemShut {NoStop}%
\bibitem [{\citenamefont {Anderson}(1959)}]{anderson59}%
  \BibitemOpen
  \bibfield  {author} {\bibinfo {author} {\bibfnamefont {P.~W.}\ \bibnamefont
  {Anderson}},\ }\bibfield  {title} {\enquote {\bibinfo {title} {{Theory of
  dirty superconductors}},}\ }\href {\doibase 10.1016/0022-3697(59)90036-8}
  {\bibfield  {journal} {\bibinfo  {journal} {J. Phys. Chem. Solids}\ }\textbf
  {\bibinfo {volume} {11}},\ \bibinfo {pages} {26} (\bibinfo {year}
  {1959})}\BibitemShut {NoStop}%
\bibitem [{\citenamefont {Abrikosov}\ and\ \citenamefont
  {Gor'kov}(1961)}]{abrikosov61}%
  \BibitemOpen
  \bibfield  {author} {\bibinfo {author} {\bibfnamefont {A.~A.}\ \bibnamefont
  {Abrikosov}}\ and\ \bibinfo {author} {\bibfnamefont {L.~P.}\ \bibnamefont
  {Gor'kov}},\ }\bibfield  {title} {\enquote {\bibinfo {title} {{Contribution
  to the theory of superconducting alloys with paramagnetic impurities}},}\
  }\href@noop {} {\bibfield  {journal} {\bibinfo  {journal} {Sov. Phys. JETP}\
  }\textbf {\bibinfo {volume} {12}},\ \bibinfo {pages} {1243} (\bibinfo {year}
  {1961})}\BibitemShut {NoStop}%
\bibitem [{\citenamefont {Kempkes}\ \emph {et~al.}(2019)\citenamefont
  {Kempkes}, \citenamefont {Slot}, \citenamefont {Freeney}, \citenamefont
  {Zevenhuizen}, \citenamefont {Vanmaekelbergh}, \citenamefont {Swart},\ and\
  \citenamefont {Smith}}]{kempkes19}%
  \BibitemOpen
  \bibfield  {author} {\bibinfo {author} {\bibfnamefont {S.~N}\ \bibnamefont
  {Kempkes}}, \bibinfo {author} {\bibfnamefont {M.~R.}\ \bibnamefont {Slot}},
  \bibinfo {author} {\bibfnamefont {S.~E.}\ \bibnamefont {Freeney}}, \bibinfo
  {author} {\bibfnamefont {S.~J.~M.}\ \bibnamefont {Zevenhuizen}}, \bibinfo
  {author} {\bibfnamefont {D.}~\bibnamefont {Vanmaekelbergh}}, \bibinfo
  {author} {\bibfnamefont {I.}~\bibnamefont {Swart}}, \ and\ \bibinfo {author}
  {\bibfnamefont {C.~Morais}\ \bibnamefont {Smith}},\ }\bibfield  {title}
  {\enquote {\bibinfo {title} {{Design and characterization of electrons in a
  fractal geometry}},}\ }\href {\doibase 10.1038/s41567-018-0328-0} {\bibfield
  {journal} {\bibinfo  {journal} {Nat. Phys.}\ }\textbf {\bibinfo {volume}
  {15}},\ \bibinfo {pages} {127} (\bibinfo {year} {2019})}\BibitemShut
  {NoStop}%
\bibitem [{\citenamefont {Zhu}\ \emph {et~al.}(2003)\citenamefont {Zhu},
  \citenamefont {Ahn}, \citenamefont {Nussinov}, \citenamefont {Lookman},
  \citenamefont {Balatsky},\ and\ \citenamefont {Bishop}}]{zhu03}%
  \BibitemOpen
  \bibfield  {author} {\bibinfo {author} {\bibfnamefont {J.-X.}\ \bibnamefont
  {Zhu}}, \bibinfo {author} {\bibfnamefont {K.~H.}\ \bibnamefont {Ahn}},
  \bibinfo {author} {\bibfnamefont {Z.}~\bibnamefont {Nussinov}}, \bibinfo
  {author} {\bibfnamefont {T.}~\bibnamefont {Lookman}}, \bibinfo {author}
  {\bibfnamefont {A.~V.}\ \bibnamefont {Balatsky}}, \ and\ \bibinfo {author}
  {\bibfnamefont {A.~R.}\ \bibnamefont {Bishop}},\ }\bibfield  {title}
  {\enquote {\bibinfo {title} {Elasticity-driven nanoscale electronic structure
  in superconductors},}\ }\href {\doibase 10.1103/PhysRevLett.91.057004}
  {\bibfield  {journal} {\bibinfo  {journal} {Phys. Rev. Lett.}\ }\textbf
  {\bibinfo {volume} {91}},\ \bibinfo {pages} {057004} (\bibinfo {year}
  {2003})}\BibitemShut {NoStop}%
\bibitem [{\citenamefont {Vojta}(2006)}]{thomas06}%
  \BibitemOpen
  \bibfield  {author} {\bibinfo {author} {\bibfnamefont {T.}~\bibnamefont
  {Vojta}},\ }\bibfield  {title} {\enquote {\bibinfo {title} {{Rare region
  effects at classical, quantum and nonequilibrium phase transitions}},}\
  }\href {\doibase 10.1088/0305-4470/39/22/R01} {\bibfield  {journal} {\bibinfo
   {journal} {J. Phys. A: Math. Gen.}\ }\textbf {\bibinfo {volume} {39}},\
  \bibinfo {pages} {R143} (\bibinfo {year} {2006})}\BibitemShut {NoStop}%
\bibitem [{\citenamefont {Jagannathan}(2004)}]{anu04}%
  \BibitemOpen
  \bibfield  {author} {\bibinfo {author} {\bibfnamefont {A.}~\bibnamefont
  {Jagannathan}},\ }\bibfield  {title} {\enquote {\bibinfo {title} {{Quantum
  Spins and Quasiperiodicity: A Real Space Renormalization Group Approach}},}\
  }\href {\doibase 10.1103/PhysRevLett.92.047202} {\bibfield  {journal}
  {\bibinfo  {journal} {Phys. Rev. Lett.}\ }\textbf {\bibinfo {volume} {92}},\
  \bibinfo {pages} {047202} (\bibinfo {year} {2004})}\BibitemShut {NoStop}%
\bibitem [{\citenamefont {Viebahn}\ \emph {et~al.}(2019)\citenamefont
  {Viebahn}, \citenamefont {Sbroscia}, \citenamefont {Carter}, \citenamefont
  {Yu},\ and\ \citenamefont {Schneider}}]{viebahn19}%
  \BibitemOpen
  \bibfield  {author} {\bibinfo {author} {\bibfnamefont {K.}~\bibnamefont
  {Viebahn}}, \bibinfo {author} {\bibfnamefont {M.}~\bibnamefont {Sbroscia}},
  \bibinfo {author} {\bibfnamefont {E.}~\bibnamefont {Carter}}, \bibinfo
  {author} {\bibfnamefont {J.C.}\ \bibnamefont {Yu}}, \ and\ \bibinfo {author}
  {\bibfnamefont {U.}~\bibnamefont {Schneider}},\ }\bibfield  {title} {\enquote
  {\bibinfo {title} {Matter-wave diffraction from a quasicrystalline optical
  lattice},}\ }\href {\doibase 10.1103/PhysRevLett.122.110404} {\bibfield
  {journal} {\bibinfo  {journal} {Phys. Rev. Lett.}\ }\textbf {\bibinfo
  {volume} {122}},\ \bibinfo {pages} {110404} (\bibinfo {year}
  {2019})}\BibitemShut {NoStop}%
\bibitem [{\citenamefont {Flicker}\ and\ \citenamefont {van
  Wezel}(2015)}]{flicker15}%
  \BibitemOpen
  \bibfield  {author} {\bibinfo {author} {\bibfnamefont {F.}~\bibnamefont
  {Flicker}}\ and\ \bibinfo {author} {\bibfnamefont {J.}~\bibnamefont {van
  Wezel}},\ }\bibfield  {title} {\enquote {\bibinfo {title} {One-dimensional
  quasicrystals from incommensurate charge order},}\ }\href {\doibase
  10.1103/PhysRevLett.115.236401} {\bibfield  {journal} {\bibinfo  {journal}
  {Phys. Rev. Lett.}\ }\textbf {\bibinfo {volume} {115}},\ \bibinfo {pages}
  {236401} (\bibinfo {year} {2015})}\BibitemShut {NoStop}%
\bibitem [{\citenamefont {Sagi}\ and\ \citenamefont {Nussinov}(2016)}]{sagi16}%
  \BibitemOpen
  \bibfield  {author} {\bibinfo {author} {\bibfnamefont {E.}~\bibnamefont
  {Sagi}}\ and\ \bibinfo {author} {\bibfnamefont {Z.}~\bibnamefont
  {Nussinov}},\ }\bibfield  {title} {\enquote {\bibinfo {title} {Emergent
  quasicrystals in strongly correlated systems},}\ }\href {\doibase
  10.1103/PhysRevB.94.035131} {\bibfield  {journal} {\bibinfo  {journal} {Phys.
  Rev. B}\ }\textbf {\bibinfo {volume} {94}},\ \bibinfo {pages} {035131}
  (\bibinfo {year} {2016})}\BibitemShut {NoStop}%
\bibitem [{\citenamefont {Wei{\ss}e}\ \emph {et~al.}(2006)\citenamefont
  {Wei{\ss}e}, \citenamefont {Wellein}, \citenamefont {Alvermann},\ and\
  \citenamefont {Fehske}}]{weisse06}%
  \BibitemOpen
  \bibfield  {author} {\bibinfo {author} {\bibfnamefont {A.}~\bibnamefont
  {Wei{\ss}e}}, \bibinfo {author} {\bibfnamefont {G.}~\bibnamefont {Wellein}},
  \bibinfo {author} {\bibfnamefont {A.}~\bibnamefont {Alvermann}}, \ and\
  \bibinfo {author} {\bibfnamefont {H.}~\bibnamefont {Fehske}},\ }\bibfield
  {title} {\enquote {\bibinfo {title} {{The kernel polynomial method}},}\
  }\href {\doibase 10.1103/RevModPhys.78.275} {\bibfield  {journal} {\bibinfo
  {journal} {Rev. Mod. Phys.}\ }\textbf {\bibinfo {volume} {78}},\ \bibinfo
  {pages} {275} (\bibinfo {year} {2006})}\BibitemShut {NoStop}%
\bibitem [{\citenamefont {Covaci}\ \emph {et~al.}(2010)\citenamefont {Covaci},
  \citenamefont {Peeters},\ and\ \citenamefont {Berciu}}]{covaci10}%
  \BibitemOpen
  \bibfield  {author} {\bibinfo {author} {\bibfnamefont {L.}~\bibnamefont
  {Covaci}}, \bibinfo {author} {\bibfnamefont {F.~M.}\ \bibnamefont {Peeters}},
  \ and\ \bibinfo {author} {\bibfnamefont {M.}~\bibnamefont {Berciu}},\
  }\bibfield  {title} {\enquote {\bibinfo {title} {{Efficient Numerical
  Approach to Inhomogeneous Superconductivity: The Chebyshev-Bogoliubov-de
  Gennes Method}},}\ }\href {\doibase 10.1103/PhysRevLett.105.167006}
  {\bibfield  {journal} {\bibinfo  {journal} {Phys. Rev. Lett.}\ }\textbf
  {\bibinfo {volume} {105}},\ \bibinfo {pages} {167006} (\bibinfo {year}
  {2010})}\BibitemShut {NoStop}%
\bibitem [{\citenamefont {Ma}\ and\ \citenamefont {Lee}(1985)}]{ma85}%
  \BibitemOpen
  \bibfield  {author} {\bibinfo {author} {\bibfnamefont {M.}~\bibnamefont
  {Ma}}\ and\ \bibinfo {author} {\bibfnamefont {P.~A.}\ \bibnamefont {Lee}},\
  }\bibfield  {title} {\enquote {\bibinfo {title} {{Localized
  superconductors}},}\ }\href {\doibase 10.1103/PhysRevB.32.5658} {\bibfield
  {journal} {\bibinfo  {journal} {Phys. Rev. B}\ }\textbf {\bibinfo {volume}
  {32}},\ \bibinfo {pages} {5658} (\bibinfo {year} {1985})}\BibitemShut
  {NoStop}%
\end{thebibliography}
\end{document}